\documentclass[floats,floatfix,showpacs,amssymb,prd,twocolumn,superscriptaddress,nofootinbib]{revtex4-1}
\newlength{\figw} 
\usepackage{graphicx,epsf, epsfig, amssymb}
\usepackage{bm}
\usepackage{longtable,tabularx}
\usepackage{color}
\usepackage[breaklinks]{hyperref}
\usepackage{amsfonts,amsmath,amssymb,mathrsfs,gensymb}
\usepackage{natbib}
\usepackage{aas_macros}
\usepackage{array}
\usepackage{multirow}
\usepackage{rotating,array}
\usepackage{graphicx}
\newcommand\unit[1]{{\rm #1}}
\newcommand\optional[1]{}
\def\be{\begin{equation}}
\def\ee{\end{equation}}
\def\beq{\begin{eqnarray}}
\def\eeq{\end{eqnarray}}
\usepackage{comment}
\newcommand{\sun}{\ensuremath{\odot}}

\begin{document}

\title{Distinguishing black-hole spin-orbit resonances by their gravitational-wave signatures}

\author{Davide Gerosa}
\email{d.gerosa@damtp.cam.ac.uk}
\affiliation{Department of Applied Mathematics and Theoretical Physics, Centre for Mathematical Sciences, University of Cambridge, Wilberforce Road, Cambridge CB3 0WA, United Kingdom}

\author{Richard O'Shaughnessy}
\email{oshaughn@gravity.phys.uwm.edu}
\affiliation{Center for Gravitation and Cosmology, University of Wisconsin-Milwaukee, Milwaukee, Wisconsin 53211, USA}

\author{Michael Kesden}
\email{kesden@utdallas.edu}
\affiliation{Department of Physics, The University of Texas at Dallas, Richardson, Texas 75080, USA }

\author{Emanuele Berti}
\email{eberti@olemiss.edu}
\affiliation{Department of Physics and Astronomy, The University of
Mississippi, University, Mississippi 38677, USA}

\author{Ulrich Sperhake}
\email{u.sperhake@damtp.cam.ac.uk}
\affiliation{Department of Applied Mathematics and Theoretical Physics, Centre for Mathematical Sciences, University of Cambridge, Wilberforce Road, Cambridge CB3 0WA, United Kingdom}
\affiliation{Department of Physics and Astronomy, The University of
Mississippi, University, Mississippi 38677, USA}
\affiliation{California Institute of Technology, Pasadena, California 91125, USA}


\date{\today}

\begin{abstract}

If binary black holes form following the successive core collapses of sufficiently massive binary stars, precessional
dynamics may align their spins $\mathbf S_1$ and $\mathbf S_2$ and the orbital angular momentum $\mathbf L$
into a plane in which they jointly precess about the total angular momentum $\mathbf J$.  These spin orientations
are known as spin-orbit resonances since $\mathbf S_1$, $\mathbf S_2$, and $\mathbf L$ all precess at the same
frequency to maintain their planar configuration.  Two families of such spin-orbit resonances exist, 
differentiated by whether the components of the two spins in the orbital plane are either aligned or antialigned.  The
fraction of binary black holes in each family is determined by the stellar evolution of their progenitors, so if
gravitational-wave detectors could measure this fraction they could provide important insights into astrophysical
formation scenarios for binary black holes.  In this paper, we show that even under the conservative assumption that
binary black holes are observed along the direction of $\mathbf J$ (where precession-induced modulations to the
gravitational waveforms are minimized), the waveforms of many members of each resonant family can be
distinguished from all members of the other family in events with signal-to-noise ratios $\rho \simeq 10$, typical of
those expected for the first detections with Advanced LIGO/Virgo. We hope that our preliminary findings inspire a
greater appreciation of the capability of gravitational-wave detectors to constrain stellar astrophysics and stimulate
further studies of the distinguishability of spin-orbit resonant families in more expanded regions of binary black-hole
parameter space.

\end{abstract}
\maketitle 

\section{Introduction}

Gravitational waves (GWs) emitted during the inspiral of binary black holes (BBHs) are expected to be an important
source \cite{2010CQGra..27q3001A} for future networks of GW detectors such the Advanced Laser Interferometer
Gravitational Wave Observatory (LIGO) and Virgo
\cite{2010CQGra..27h4006H}, LIGO-India \cite{2013IJMPD..2241010U}, the  Kamioka Gravitational Wave Detector (KAGRA) \cite{2012CQGra..29l4007S}, and the Einstein Telescope \cite{2010CQGra..27s4002P}. 
These BBHs can form in two distinct channels: (1) mass segregation can cause isolated black holes (BHs) to sink to the centers of
dense stellar clusters and dynamically form binaries \cite{2006ApJ...637..937O,2007PhRvD..76f1504O}, or (2)
massive binary stars can evolve into BBHs if each member of the binary is sufficiently massive at the time of core
collapse and binary evolution does not destroy the binary before both stars have had the chance to collapse into BHs
\cite{2012ApJ...759...52D,2013ApJ...779...72D}.  Once formed, BBHs emit GWs that extract energy and angular
momentum from the orbit, decreasing the binary separation and increasing the orbital frequency (and thus the GW
frequency). Most binaries are expected to circularize by the time they enter the sensitivity band of ground-based detectors \cite{1963PhRv..131..435P,1964PhRv..136.1224P} (see \cite{2013PhRvD..87d3004E} and references therein for recent work on eccentric binary rates and detection strategies). Circular BBH inspirals are characterized by eight intrinsic parameters: the masses $m_1$ and $m_2$ of each BH and their spins
$\mathbf S_1$ and $\mathbf S_2$.  We choose without loss of generality for the first BH to be more massive than the
second: $m_1 > m_2$.  The spectrum of emitted GWs depends on these eight parameters, which can
therefore be measured by GW detectors if the sources are observed with a sufficient signal-to-noise ratio $\rho$.  The
distributions of these intrinsic parameters depend on how the BBHs form, allowing GW parameter estimation to
constrain not just individual BBH systems, but their astrophysical formation channels as well.

\begin{figure}
\centering
\includegraphics[width=0.9\columnwidth]{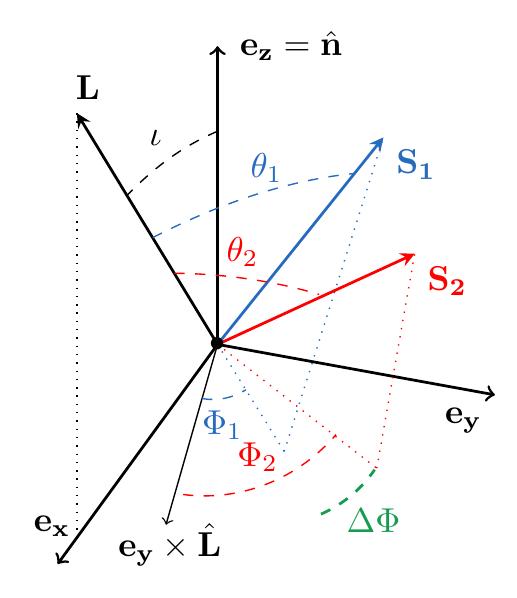}
\caption{(color online) Conventions and definitions used in this paper. We work in the radiation frame, where the $z$
axis is oriented along the line of sight $\hat{\mathbf{n}}$. The orbital angular momentum $\mathbf{L}$ lies in the
$xz$-plane at $f_{\rm ref}$ and is inclined by an angle $\iota$ with respect to the line of sight. The directions of the spins
$\mathbf{S_1}$ (blue) and $\mathbf{S_2}$ (red) are specified using polar
angles $\theta_i$ and azimuthal angles $\Phi_i$ $(i=1,\,2)$, which are defined in a frame where the $z$ axis is aligned
with the orbital angular momentum $\mathbf{L}$.  As resonant binaries precess, their orbital angular momentum and
spins remain coplanar implying that the angle $\Delta\Phi=\Phi_2 - \Phi_1$ (green) remains fixed at either $0^\circ$ or
$\pm180^\circ$.  In later sections of the paper, we will fix $\iota$ and $\Phi_1$ by aligning the line of sight with the total angular momentum: cf.  Eqs.~(\ref{Jalongn_phi1}) and (\ref{Jalongn}). 
}
\label{radframe}
\end{figure}

Our focus in this paper is on whether BBH spin orientations can be measured with sufficient accuracy in $\rho \simeq
10$ sources to constrain the formation of binaries.  BBH spin directions are described by three parameters: the two
angles $\theta_i$ between spins $\mathbf S_i$ and the orbital angular momentum $\mathbf L$ and the angle
$\Delta\Phi = \Phi_2 - \Phi_1$ between the components of the two spins in the orbital plane (see Fig.~\ref{radframe}).  Although the individual
angles $\Phi_i$ of each of the BBH spins are among the 8 observable intrinsic parameters listed above, only their
difference $\Delta\Phi$ provides constraints on BBH formation in the absence of an additional intrinsic vector to break
the axisymmetry of the equatorial plane.  In the first astrophysical formation
channel described above, the BBHs form independently and the dynamical formation of the binary should not
depend on the BH spin.  We therefore expect both BH spins to have isotropic orientations, in which case the
post-Newtonian (PN) GW inspiral will preserve the isotropy of the BBH spins \cite{2007ApJ...661L.147B}.  This is not
the case, however, in the second astrophysical formation channel, where the BBHs inherit the directions of their spins
from their stellar progenitors.

In a previous paper \cite{2013PhRvD..87j4028G}, we examined how the spins of BBHs formed from stellar binaries
depend on the evolution of their stellar progenitors.  Throughout this evolution, the initially more massive star will be
designated  as the ``primary" and the less massive star will be called the ``secondary".  The binary evolution proceeds in
several stages:
\begin{itemize}

\item[(a)] The binary stars initially have spins aligned with their orbital angular momentum $\mathbf L$ as tidal
alignment occurs on a much shorter timescale than the main-sequence lifetimes of the stars \cite{2001ApJ...562.1012E}.

\item[(b)] The more massive primary evolves more quickly than the secondary, filling its Roche lobe and transferring
mass to the secondary.

\item[(c)] The core of the primary collapses, forming a BH and detonating a supernova explosion.  This asymmetric
explosion kicks the binary and tilts the orbital plane.  The directions of the stellar spins remain unchanged and thus
become misaligned with the new direction of $\mathbf L$.

\item[(d)] Tides align the spin of the secondary with the new direction of $\mathbf L$ while leaving the spin of the more
compact BH unchanged.

\item[(e)] The core of the secondary collapses into a BH.  The orbital plane is tilted a second time, misaligning the spin
of the secondary with the new direction of $\mathbf L$ and on average increasing the misalignment of the spin of the
primary even further.

\item[(f)] The BBH spins precess many times before the frequency of emitted GWs enters the sensitivity band of
ground-based detectors.

\end{itemize}
Although the PN spin precession in stage (f) above leaves isotropic spin distributions isotropic as the BBHs inspiral,
it can profoundly affect anisotropic spin distributions resulting from stages (a) - (e).

The manner in which spin precession alters the distribution of BBH spins can best be understood by appreciating the
influence of PN spin-orbit resonances, first identified by Schnittman \cite{2004PhRvD..70l4020S}.  BBHs evolve on
three distinct timescales: (1) the orbital time $t_{\rm orb} \sim (r^3/GM)^{1/2}$, (2) the precession time $t_{\rm pre}
\sim c^2r^{5/2}/[\eta(GM)^{3/2}] \sim (t_{\rm orb}/\eta)(r/r_g)$, and (3) the radiation-reaction time $t_{\rm RR} \sim
E/|dE_{\rm GW}/dt| \sim c^5 r^4/[\eta(GM)^3] \sim (t_{\rm orb}/\eta)(r/r_g)^{5/2}$, where $M = m_1 + m_2$ is the total
mass, $\eta = m_1m_2/M^2$ is the symmetric mass ratio, and $r_g = GM/c^2$ is the gravitational radius.  In the PN
regime, $r \gg r_g$ and these timescales are widely separated: $t_{\rm orb} \ll t_{\rm pre} \ll t_{\rm RR}$.  In this limit,
we can average the spin-precession equations
\cite{1995PhRvD..52..821K,2009PhRvD..79j4023A,2011PhRvD..84d9901A} over an orbit while leaving the total
angular momentum $\mathbf{J} = \mathbf{L} + \mathbf{S_1} + \mathbf{S_2}$ fixed.  The three angular momenta
$\mathbf L$, $\mathbf S_1$, and $\mathbf S_2$ will generally span three-dimensional space at any given time and
precess in a complicated fashion on the precession time $t_{\rm pre}$ that preserves the magnitude and direction of 
$\mathbf J$.  However, Schnittman discovered special spin configurations in which $\mathbf L$, $\mathbf S_1$, and
$\mathbf S_2$ would remain in a two-dimensional plane and jointly precess about $\mathbf J$ on the precession
time $t_{\rm pre}$ \cite{2004PhRvD..70l4020S}.  He called these configurations ``spin-orbit resonances'' because
$\mathbf L$, $\mathbf S_1$, and $\mathbf S_2$ all precessed about $\mathbf J$ at the same frequency.

These spin-orbit resonances are divided into two families: resonances in which the spin components in the orbital plane
are aligned ($\Delta\Phi = 0^\circ$) and those in which these components are antialigned ($\Delta\Phi = \pm 180^\circ$).
At a given binary separation $r$ [or GW frequency $f = \pi^{-1}(GM/r^3)^{1/2}$], each of the two resonant families defines a
different curve in the $\theta_1\theta_2$-plane.  As resonant BBHs inspiral on the longer radiation-reaction time
$t_{\rm RR}$, they remain in spin-orbit resonances though the values of $\theta_i$ vary as the relationship between
$\theta_1$ and $\theta_2$ for the spin-orbit resonances is a function of the separation $r$.  

One might imagine that since the one-parameter spin-orbit resonances constitute a set of measure zero in the
three-dimensional parameter space ($\theta_1$, $\theta_2$, $\Delta\Phi$) of spin configurations at any given
separation, they are merely a mathematical curiosity of little relevance to astrophysical BBHs.  However, BBHs near
a spin-orbit resonance will be influenced by its presence, with $\Delta\Phi$ librating about $0^\circ$ or
$\pm 180^\circ$ rather than circulating through the full range $\Delta\Phi \in [-180^\circ, +180^\circ]$.  Furthermore,
as the binary separation decreases an increasing fraction of BBHs will be captured into this librating portion of the
parameter space.  Which of the two families will be favored by this capture process, the $\Delta\Phi = 0^\circ$
resonances or the $\Delta\Phi = \pm 180^\circ$ resonances?  The answer to this question depends on the distribution
of $\theta_i$ at large separations.  BBHs where the spin of the more massive BH is less misaligned with the orbital
angular momentum than that of the less massive BH ($\theta_1 < \theta_2$) will be preferentially attracted to the 
$\Delta\Phi = 0^\circ$ family of resonances, while BBHs for which $\theta_1 > \theta_2$ will be preferentially attracted
to the  $\Delta\Phi = \pm 180^\circ$ family \cite{2004PhRvD..70l4020S}.  

The distribution of $\theta_i$ at large separations is determined by the astrophysics of BBH formation.   If the tidal
alignment of the secondary's spin in stage (d) above is efficient, the primary's spin will on average be more
misaligned with the orbital angular momentum than the secondary's at the start of PN spin precession in stage (f),
since its misalignment will have been built up in {\it both} supernova recoils in stages (c) and (e)\footnote{
The second kick is more likely to increase the misalignment between the orbital angular momentum and the spin of the
primary because of the greater amount of phase space at larger values of $\theta$ (the Jacobian determinant
$\sin\theta$ increases with $\theta$) provided the first tilt is $\lesssim\pi/2$. See \cite{2013PhRvD..87j4028G} for a
discussion.}
However, the 
primary star (which is initially more massive and is thus first to collapse into a BH) will not always become the more massive
BH.  If enough mass is transferred from the primary to the secondary prior to the first core collapse in stage (b), the
primary will evolve into the less massive BH.  We will refer to this possibility as the reverse-mass-ratio (RMR) scenario
\cite{2013PhRvD..87j4028G}.  In this case, the more massive BH (evolved from the secondary) will have a less
misaligned spin ($\theta_1 < \theta_2$), and the $\Delta\Phi = 0^\circ$ family of resonances will be preferentially
populated.  Conversely, in the standard-mass-ratio (SMR) scenario where the primary evolves into the more massive
BH, it will have a more misaligned spin ($\theta_1 > \theta_2$) and the $\Delta\Phi = \pm 180^\circ$ family of
resonances will be favored.  In the ``No Tides" scenario where the tidal alignment in stage (d) is ineffective, neither of the
resonant families will be favored over the other.

Our previous paper \cite{2013PhRvD..87j4028G} showed that for a simplified but not unreasonable toy model of the
BBH formation described above, a large fraction of BBHs were librating about the $\Delta\Phi = 0^\circ~(\pm
180^\circ)$ resonances in the RMR (SMR) scenario by the time the GW frequency $f$ approached the frequency at which
most of the signal-to-noise ratio (SNR) is accumulated ($\sim 60$ Hz).  In this paper, we investigate what SNR $\rho$ is required
to distinguish the GWs emitted by BBHs in the two resonant families.  A thorough exploration of the full intrinsic and
extrinsic parameter space that characterizes BBH waveforms is computationally prohibitive, so we make several mostly
conservative assumptions to restrict this parameter space.  To facilitate comparison with our previous paper, we fix
$m_1 = 7.5~M_\odot$, $m_2 = 6~M_\odot$, and $\chi_i  \equiv S_i/m_i^2 = 1$.  These values are close to the expected
peak of the distribution of astrophysical BH binaries detectable by Advanced LIGO, as predicted by population-synthesis
codes \cite{2012ApJ...759...52D}.  They are also consistent with the strong influence of PN precession, as binaries are most
effectively captured into spin-orbit resonances when the BBH masses are comparable ($q \equiv m_2/m_1 \lesssim 1$)
and both dimensionless spin amplitudes are large ($\chi_i \gtrsim 0.5$)
\cite{2010PhRvD..81h4054K,2013PhRvD..87j4028G}.  All BBHs, therefore, have the same masses and spin magnitudes,
ensuring that the spin directions are solely responsible for the differences in the waveforms.  We also choose the
position $\hat{\mathbf n}$ of the BBHs on the sky such that they are directly overhead of the GW detectors. We align
the direction $\hat{\mathbf J}$ of the total angular momentum with $\hat{\mathbf n}$ at a reference frequency
$f_{\rm ref} = 60$~Hz, a typical frequency at which most of the SNR is accumulated.  This latter choice is conservative since
$\hat{\mathbf L}$ precesses about the nearly constant $\hat{\mathbf J}$ during the inspiral, and thus the precessional
modulations to the waveform due to changes in the angle between $\hat{\mathbf L}$ and $\hat{\mathbf n}$ are
minimized.  With these choices, we compare the waveforms of each member of the two families of spin-orbit
resonances with those of all of the members of the opposite family by computing their overlap ${\cal O}$.  If this overlap
with all members of the opposite family is sufficiently less than unity, we can safely claim to have determined to which of
the resonant families the BBH belongs.

The remainder of this paper is organized as follows.  
In Sec.~\ref{sec:Dynamics} we review the dynamics of resonant BBHs, introduce a convenient parametrization to
identify members of each resonant family, and show qualitatively why the two families are dynamically distinguishable. 
In Sec.~\ref{sec:Compare}, we examine the GWs emitted by resonant BBHs, use the overlap between waveforms
from different families to assess their distinguishability, then investigate how this distinguishability can be used to
differentiate between astrophysical scenarios of BBH formation.
In Sec.~\ref{sec:SingleSpin} we hypothesize that the dynamics and waveforms of resonant binaries are similar to
binaries with a single effective spin, then use this hypothesis to develop two different predictions for the best matching
waveforms in the different resonant families.
Some final remarks are provided in Sec.~\ref{S:disc}.  Some technical details concerning the numerical evolution of
the BBHs and the nature of the correspondence between matching waveforms in the two families are given in
Appendixes \ref{app_coord} and \ref{ap:Symmetry}.

Throughout the rest of this paper we use geometrical units where $G=c=1,$ and we use hats to identify unit vectors.  For
example, the direction of the orbital angular momentum will be denoted by $\hat{\mathbf L}={\mathbf L}/|{\mathbf L}|$.

\section{Review of spin-orbit resonances}
\label{sec:Dynamics}

While the dynamics of PN spin-orbit resonances has been discussed at length elsewhere \cite{2004PhRvD..70l4020S,2010PhRvD..81h4054K,2010ApJ...715.1006K,2012PhRvD..85l4049B,2013PhRvD..87j4028G},
in this section we make two new observations about the two resonant families: (1) they share a common
natural parametrization in terms of their effective spin, but (2) their qualitatively different orientations lead to
observationally distinct degrees of orbital-plane precession.  To substantiate these claims, as well as to perform all
subsequent calculations in this work,  we evolve quasicircular inspiral orbits and generate the associated
PN waveforms using the {\sc lalsuite} SpinTaylorT4 code, developed by the LIGO Collaboration \cite{lal} and
based on prior work \cite{2003PhRvD..67j4025B,2009PhRvD..80h4043B} on quasicircular spinning BH
binaries\footnote{Recent publications provide other approximations to two-spin dynamics, valid for a limited range of
  spins \cite{2013PhRvD..88l4015K,2014arXiv1404.3180C}. We
  intentionally adopt a well-tested approximation valid for generic spins.}.
We specify all binary parameters at a GW frequency $f_{\rm ref}=60$~Hz, near the peak of the squared SNR per unit frequency for advanced GW
detectors; more details on the specification of binary parameters and our calculation of waveforms are provided in
Appendix~\ref{app_coord}.

\subsection{Parametrizing spin-orbit resonances}
\label{sortingresonances}

As described in the introduction, spin-orbit resonances are solutions of the PN 
spin-precession equations
\cite{1995PhRvD..52..821K,2004PhRvD..70l4020S} for which $\mathbf L$, $\mathbf S_1$, and $\mathbf S_2$
remain coplanar throughout the inspiral.  At a given binary separation $r$ (or GW frequency $f$), there are two
different one-parameter families of spin-orbit resonances, corresponding to whether the spin components in the
orbital plane are aligned ($\Delta\Phi=0^\circ$) or antialigned ($\Delta\Phi= \pm 180^\circ$)
\cite{2004PhRvD..70l4020S,2010PhRvD..81h4054K}.  In previous work, spin-orbit resonances were identified by
their values of $\theta_1$ and $\theta_2$.  Although these angles remain constant on the precession time
$t_{\rm pre}$ for resonant binaries, they evolve on the longer radiation-reaction time $t_{\rm RR}$, and are thus less
useful for identifying BBHs as they inspiral to smaller separations.  However, if the individual BBH spins
are combined into an ``effective" spin\footnote{In his study of the
overlaps of aligned-spin BBH waveforms, Ajith considered a slightly different effective spin derived from the combination
of spins appearing in the dominant (spin-orbit) term of the GW phase and amplitude: cf. Eq.~(5.9) of \cite{2011PhRvD..84h4037A}. The two effective spins agree in the limit of small binary mass ratios.} \cite{2001PhRvD..64l4013D}
\begin{align} \label{E:S_0}
  \mathbf{S_0}= (1+q) \mathbf{S_1} + \left(1+\frac{1}{q}\right) \mathbf{S_2} ~,
\end{align}
the projection
\begin{align}
  \xi\equiv \frac{\mathbf{S_0}\cdot \hat{\mathbf L}}{M^2}
    \Bigg|_{f=f_{\rm ref}} = \frac{\chi_1 \cos\theta_1 + q \chi_2 \cos\theta_2}{1+q}
  \label{eq:defxi}
\end{align}
of this effective spin onto the orbital angular momentum is approximately conserved by orbital evolution when all known PN orders
are included \cite{2010PhRvD..81h4054K}, and is exactly conserved up to 2PN when 2PN-order radiation-reaction is used \cite{2008PhRvD..78d4021R}.  This suggests that $\xi$ can be used to parametrize
the members of each family of spin-orbit resonances throughout their inspiral.

\begin{figure}
\includegraphics[width=\columnwidth]{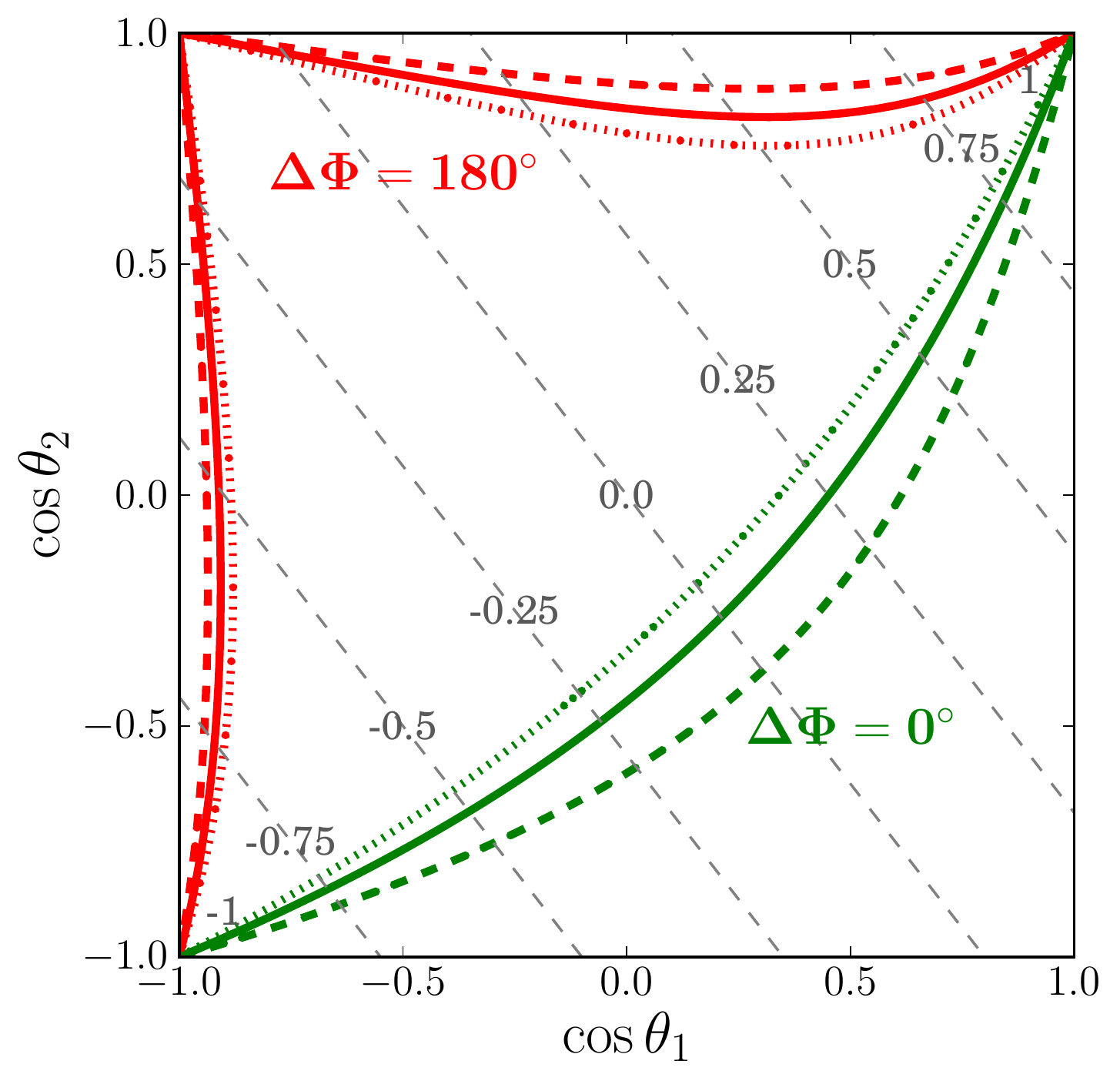}
\caption{(color online) One-parameter families of resonant binaries superimposed on contours of constant $\xi=\mathbf{S_0}\cdot
\hat{\mathbf L}/M^2$. Red (top-left) and green (bottom-right) curves
show resonant configurations in the two coplanar families for
our canonical choice of the parameters ($q=0.8$, $M=13.5 M_{\sun}$,
$\chi_1=\chi_2=1$) at three different emitted frequencies: 20 Hz (dashed),
60 Hz (i.e. $f_{\rm ref}$, solid) and $150$ Hz (dotted). The value of
$\xi\in[-1,1]$ is constant over the sloped dashed lines. Each of them
always crosses the resonant curves exactly once, thus unambiguously identifying
a single binary [i.e. a pair $(\theta_1,\theta_2)$] in each family.}
\label{res_S0L}
\end{figure}

Figure~\ref{res_S0L} shows contours of constant $\xi$ (straight dashed lines in the $\cos\theta_1\cos\theta_2$-plane) superimposed on
the colored curves corresponding to the spin-orbit resonances at different GW frequencies.  Note that
\begin{align}
  \xi \in \left[
    -\frac{\chi_1+q\chi_2}{1+q},\frac{\chi_1+q\chi_2}{1+q} \right]\,,
  \label{eq:xirange}
\end{align}
which for our choice of maximal spins $\chi_i=1$ implies $\xi \in [-1, +1]$.  Each line of constant $\xi$ crosses the
curve for each resonant family (i.e., $\Delta\Phi=0^\circ$ or $\pm180^\circ$) only once at a given GW frequency,
establishing a one-to-one correspondence between $\xi$ and resonant solutions with fixed $(\theta_1,\theta_2)$ of the PN
equations.  As resonant BBHs inspiral, they remain resonant and move towards the diagonal $\cos\theta_1 = \cos
\theta_2$ along lines of constant $\xi$.

The approximate conservation of the parameter $\xi$ suggests that resonant binaries might dynamically resemble
single-spin binaries.  We will return to the similarity between resonant and single-spin binaries in Sec.~\ref{sec:SingleSpin}.

\subsection{Different spin orientations imply different precessional dynamics}
\label{SS:diffdyn}

BBH spins are oriented in qualitatively different ways in the two families of spin-orbit resonances.  In the
common plane in which $\mathbf L$, $\mathbf S_1$, and $\mathbf S_2$ all precess, the two spins are on the same
side of the orbital angular momentum for the $\Delta \Phi=0^\circ$ resonances, and on opposite sides for the $\Delta
\Phi=\pm180^\circ$ resonances.  This implies that for comparable-mass binaries ($q \lesssim 1$) with similar spin
magnitudes ($\chi_1 \simeq \chi_2$), the component of the total spin ${\mathbf S} = {\mathbf S_1} + {\mathbf S_2}$ in
the orbital plane will be much larger for the $\Delta \Phi=0^\circ$ resonances (where the individual components in the
plane add constructively) and smaller for the $\Delta \Phi=180^\circ$ resonances (where they add destructively).  
Since the total angular momentum $\mathbf J=\mathbf L+\mathbf S$, $\mathbf J$ and $\mathbf L$ will be significantly more
misaligned for the $\Delta \Phi=0^\circ$ resonances than for the $\Delta \Phi=\pm180^\circ$ resonances.  Misalignment
between $\mathbf J$ and $\mathbf L$ leads to orbital-plane precession, which leaves an observational signature in the
gravitational waveform, as will be seen in the next section.

\begin{figure}
\includegraphics[page=1,width=\columnwidth]{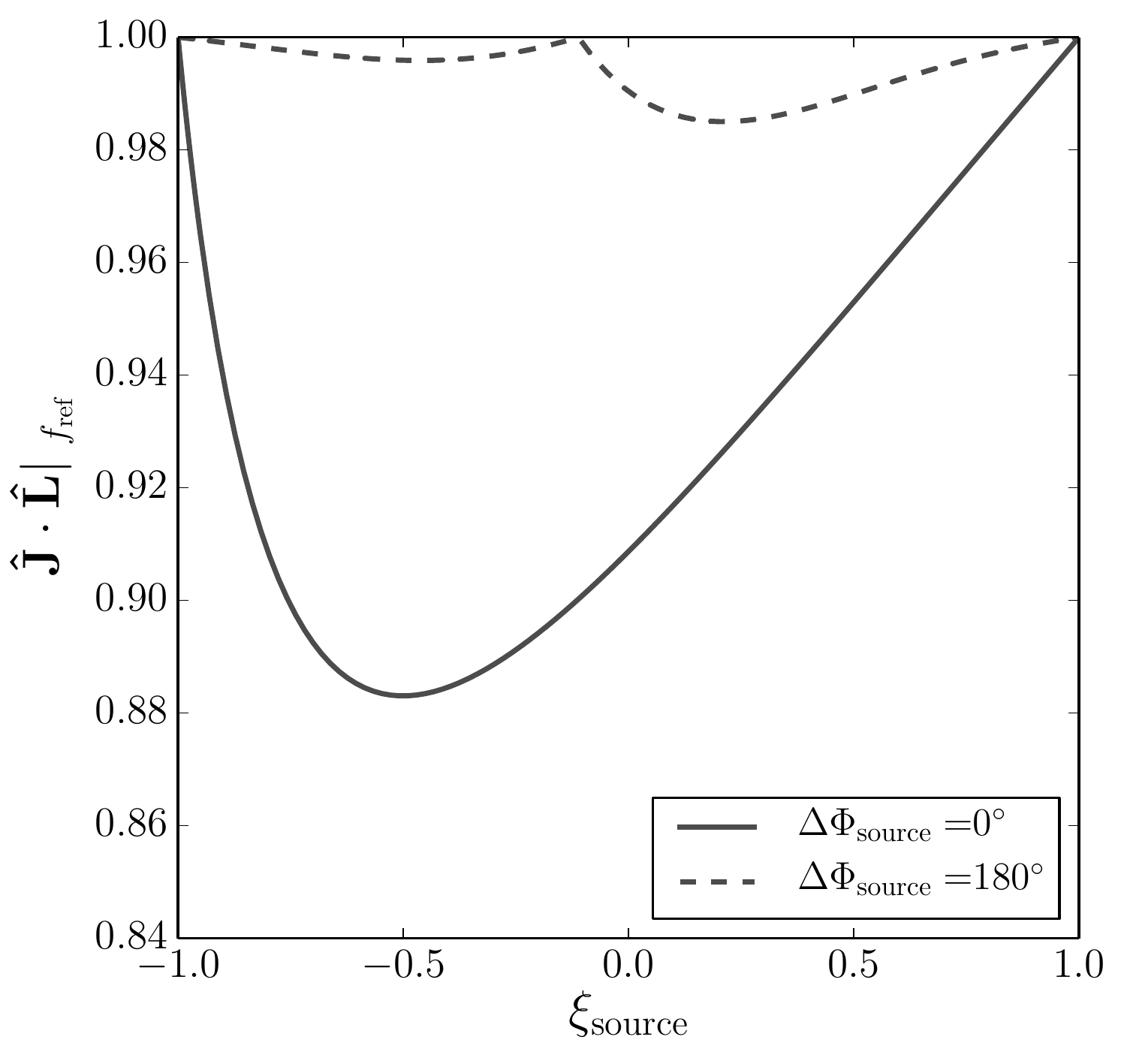}
\caption{Inner product of unit vectors in the directions of the orbital angular momentum $\mathbf L$ and total angular
momentum $\mathbf J$ for BBHs in spin-orbit resonances at the reference frequency  $f_{\rm ref}$.  Solid and dashed
curves correspond to the $\Delta\Phi=0^\circ$ and $\Delta\Phi=\pm180^\circ$ families respectively.  If the line of sight
points along $\hat{\mathbf J}$, $\hat{\mathbf J}\cdot \hat{\mathbf L}=\cos\iota$ as given by Eq.(\ref{Jalongn}).}
\label{Jn_iota}
\end{figure}

This misalignment is illustrated in Fig.~\ref{Jn_iota}, which shows the inner product of the unit orbital and total angular
momentum vectors  $\hat{\mathbf L}$ and $\hat{\mathbf J}$ as a function of effective projected spin $\xi$ for members
of the two resonant families at the reference frequency $f_{\rm ref}$.  For $\Delta \Phi = \pm180^\circ$, $\hat{\mathbf L}$
and $\hat{\mathbf J}$ are almost completely aligned for all values of $\xi$ while BBHs in the $\Delta \Phi=0^\circ$
family show significant misalignment of $\hat{\mathbf L}$ and $\hat{\mathbf J}$ unless $\xi$ is close to $\pm 1$.  Note
that $\xi=\pm 1$ corresponds to the bottom-left and upper right corners of the $\cos \theta_1$ vs $\cos \theta_2$
plane in Fig.~\ref{res_S0L}, where the two resonant families meet.  Binaries with $\xi=\pm 1$ have spins totally
aligned or antialigned with $\hat{\mathbf L}$, and therefore belong to both families.

\section{Computing and Comparing waveforms for resonant binaries}
\label{sec:Compare}

Our main goal in this paper is to compare sets of (simulated) source waveforms $h_{ab,0}(t)$ against template
waveforms $h_{ab}(t)$, where $h_{ab}$ is the transverse-traceless GW strain tensor.  
We adopt the same signal model used in \cite{2014PhRvD..89f4048O}.  Further details are provided in Appendix~\ref{app_coord}.

The incident strain $h_{ab}$ induces a linear response $h(t)$ in the GW detector that is fully characterized by two
detector-response functions $F_{+,\times}$:
\begin{eqnarray}
h(t) &=F_{+} h_+(t-{\mathbf x}\cdot \hat{\mathbf k}) + F_\times h_\times(t-{\mathbf x} \cdot \hat{\mathbf k}) 
\end{eqnarray}
where ${\mathbf x}$ is the location of the detector, $h_{+} = h^{ab}e_{ab,+}/2$, $h_\times = h^{ab}e_{ab,\times}/2$,
and $e_{+,\times}$ are basis tensors for the two-dimensional space of transverse-traceless tensors propagating in
the $\hat{\mathbf k}$ direction.  We will assume throughout this paper that the source is directly overhead of a single detector
that is oriented such that $F_+=1$ and $F_\times=0$.\footnote{\label{foot:Wolog}Without loss of generality, any
observations undertaken by a single detector can always be rescaled to be directly overhead, simply by rescaling 
$F_{+,\times}$; see, e.g., Eq. (11) of \cite{2012PhRvD..86f4020B}.}
To quantify the difference between the responses $h_0(t)$ and $h(t)$ to the source and template, respectively, we introduce the
noise-weighted inner product
\cite{1994PhRvD..49.2658C}
\begin{align}
(h_0 | h) \equiv 4 \,{\rm Re} \int_0^\infty \frac{\tilde h_0 (f) \tilde h^*(f)}{S_n(f)} df\,,
\label{innerproduct}
\end{align}
where $\tilde h(f)$ and $\tilde h_0(f)$ are the Fourier transforms of $h(t)$ and $h_0(t)$, and $S_n(f)$ is
the noise power spectral density (PSD) for a given detector/network. The
SNR $\rho$ of the source is given by
\begin{align}
\rho \equiv (h_0|h_0)^{1/2}\,,
\label{snrdef}
\end{align}
and can be evaluated by integrating the squared SNR per unit frequency
\begin{align}
\frac{d\rho^2}{df} \equiv 4  \,\frac{ | \tilde{h}_0 (f)|^2}{S_n(f)}~.
\label{drho2df}
\end{align}
The key quantity needed to perform comparisons between the source and template waveforms is the \emph{overlap}
\begin{align}
{\cal O}(h_0, h) \equiv 
\max_{t_c, \phi_c} \frac{(h_0 | h)}{\sqrt{(h_0 | h_0) (h | h)}}~,
\label{overlapdefinition}
\end{align}
the normalized inner product of the two responses maximized over the arrival time $t_c$ and the phase at
coalescence $\phi_c$.  This overlap can be used to make rigorously defined statistical statements about the
difference between two distinct waveforms and waveform families in the presence of detector noise
\cite{1994PhRvD..49.2658C,2008PhRvD..78l4020L,2013PhRvD..87b4004C}. In particular, two waveforms can be
distinguished when their difference is sufficiently large [i.e. $(h-h_0,h-h_0)\gtrsim 1$] and hence when their overlap is sufficiently small; this condition is approximately 
\begin{eqnarray}
{\cal O}(h_0, h) \lesssim 1 - \rho^{-2}~.
\label{eq:distinguish}
\end{eqnarray}

\begin{figure*}
\includegraphics[width=\textwidth]{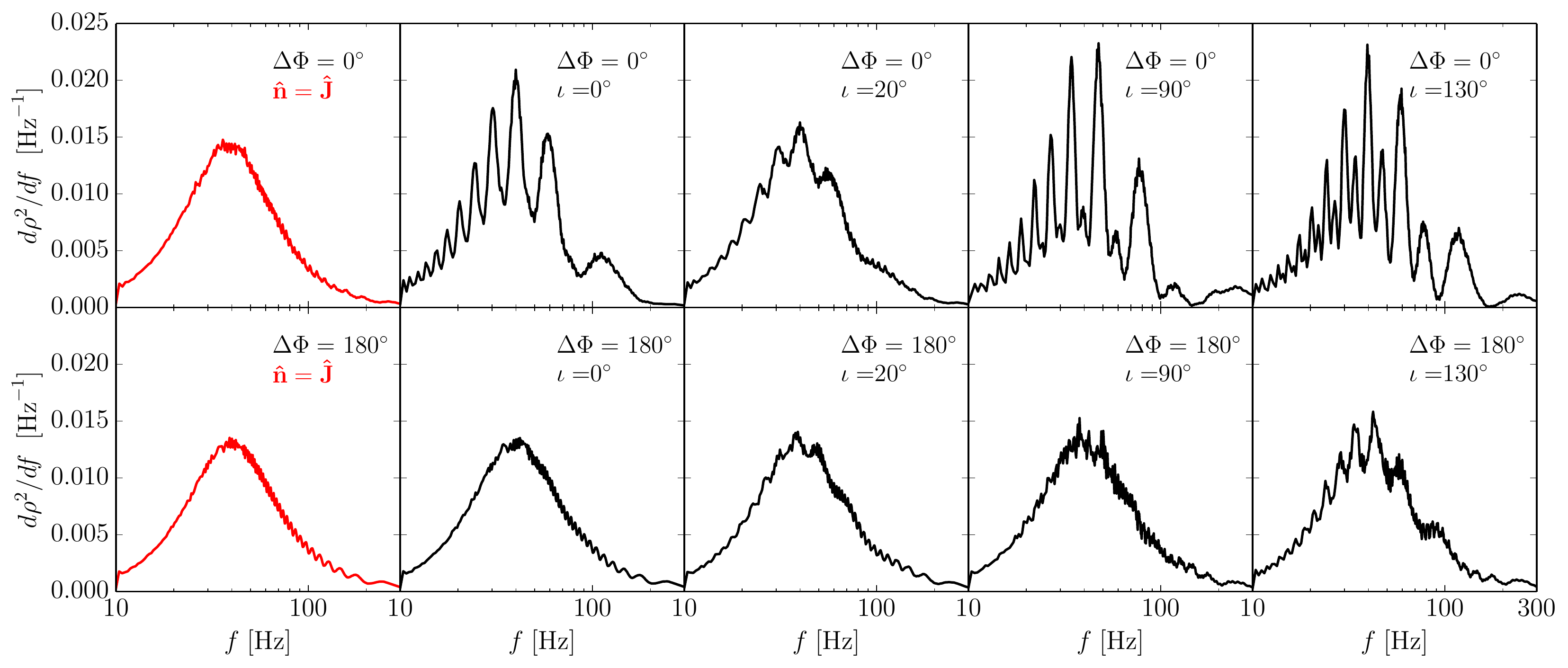}
\caption{(color online) Expected squared SNR per unit frequency [Eq.~(\ref{drho2df})] for binaries belonging to the two resonant
families.  The sources all have a projected effective spin $\xi=-0.5$, but they are viewed at different inclinations $\iota$.
Waveforms from binaries in the $\Delta\Phi=0^\circ$ resonance (top row) exhibit wider modulations due to greater
precession of the orbital plane. On the other hand, in the $\Delta\Phi=180^\circ$ family (bottom row), the components of
the two spins in the orbital plane partially cancel each other, reducing the precession of $\mathbf{L}$.  The expected
modulation varies with $\iota$ and is minimized by the values of $\iota$ predicted by Eq.(\ref{Jalongn}), for which the line
of sight $\hat{\mathbf n}$ is parallel to the total angular momentum $\mathbf{J}$ (first column, red curves).  With our
canonical choice of the parameters and $\xi=-0.5$, the $\hat{\mathbf n} = \hat{\mathbf J}$ case corresponds to
$\iota\simeq 27^\circ$ for $\Delta\Phi=0^\circ$, and $\iota\simeq 5^\circ$ for $\Delta\Phi=180^\circ$.}
\label{fig:expSNR}
\end{figure*}

\begin{figure*}
\includegraphics[width=\textwidth]{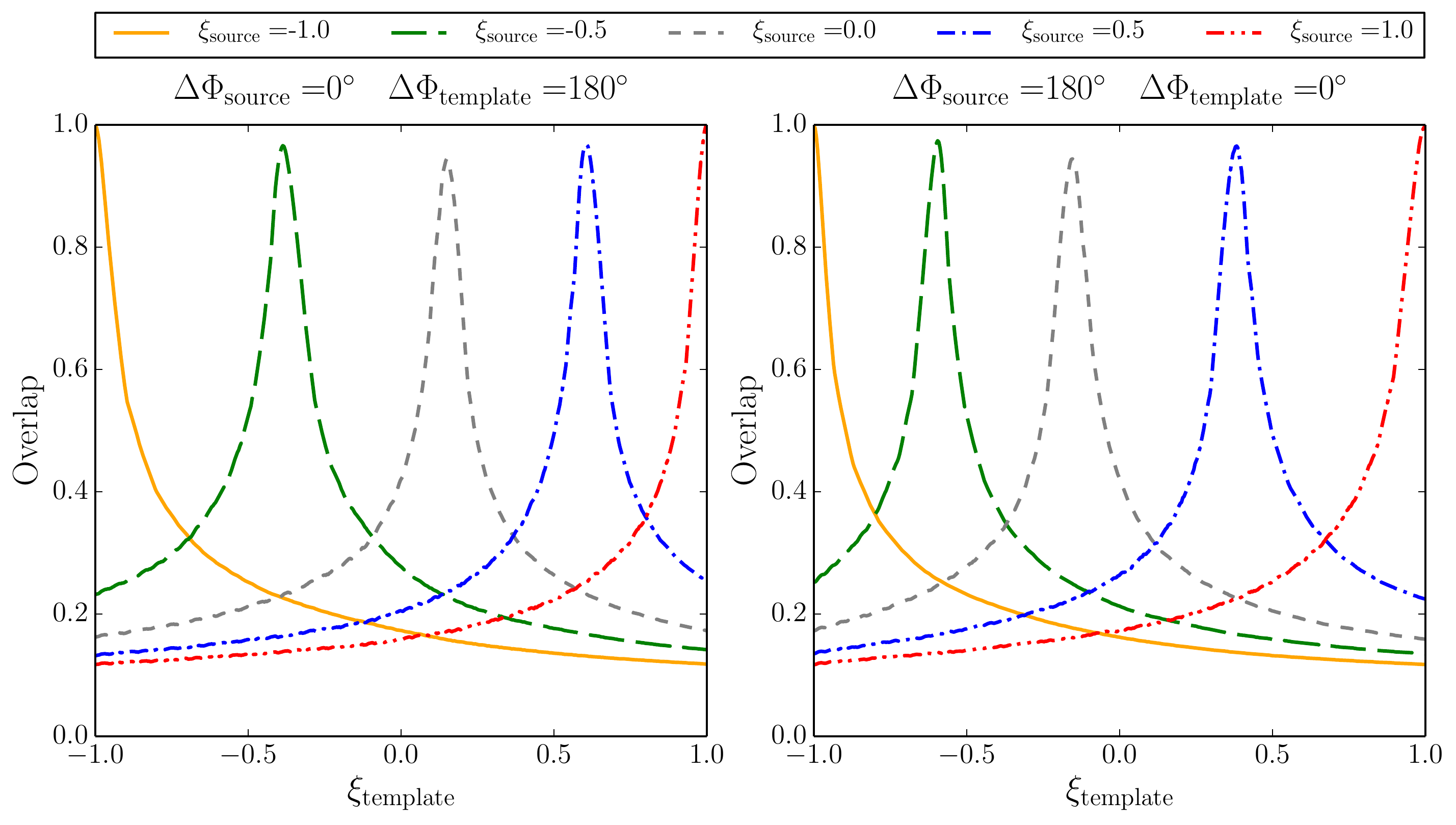}
\caption{(color online) Overlap between resonant binaries. A source in either the $\Delta\Phi=0^{\circ}$ resonance (left) or the
$\Delta\Phi=180^{\circ}$ resonance (right), is compared with members of the other family parametrized by
$\xi_{\rm template}$.  Each one-parameter family is built varying over the spin direction through $\xi$, while all the
remaining parameters are fixed.  Five different sources are considered, but the same trend holds for every value of
$\xi_{\rm source}$.  Each curve possesses a clear unique maximum, pairing the source binary with a best matching template in the other family. }
\label{xi_ov_notmax}
\end{figure*}

\begin{figure}
\includegraphics[width=\columnwidth]{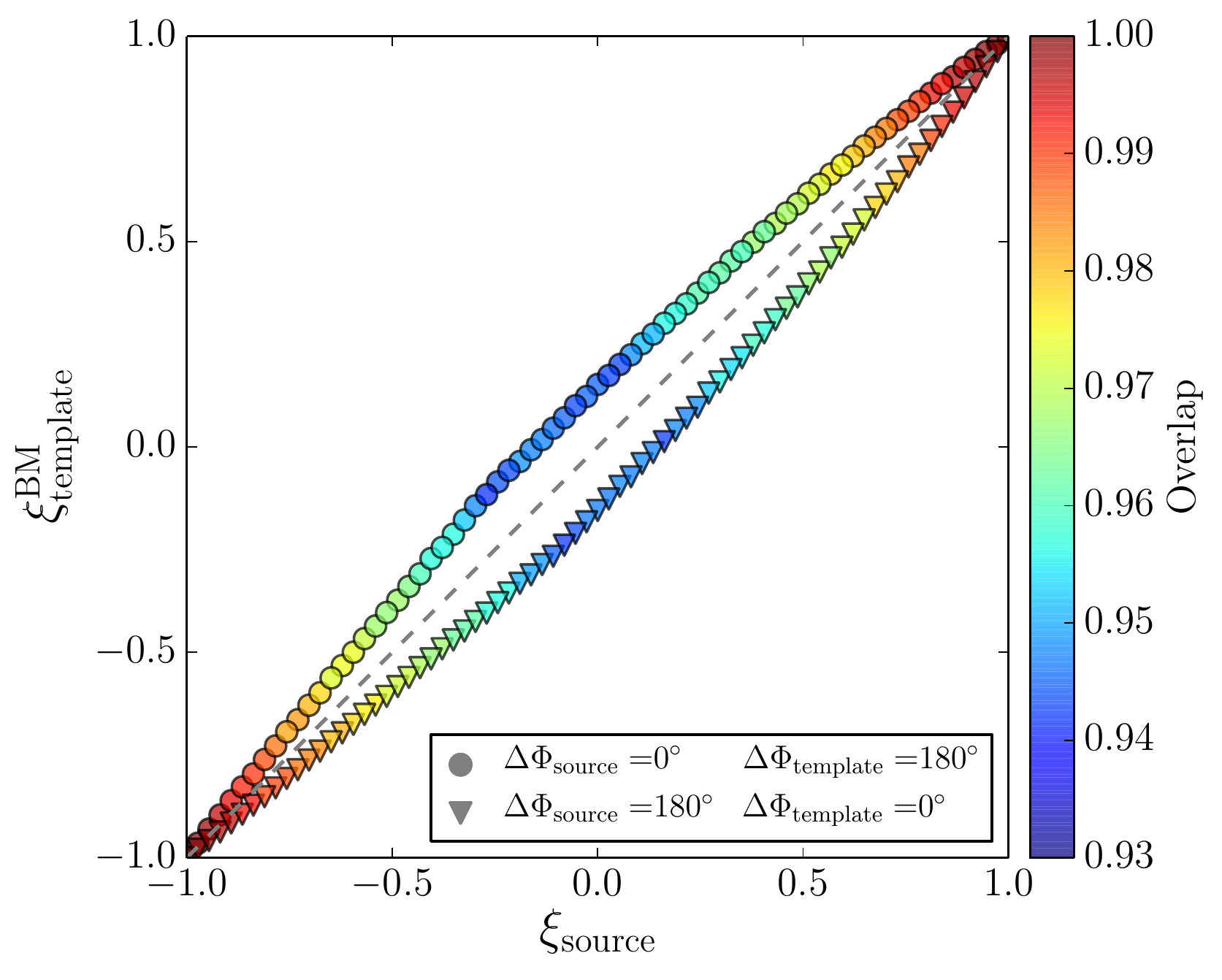}
\caption{(color online) Pairing between resonant binaries in the resonant families. Each source $\xi_{\rm source}$ is paired
with the best matching template $\xi^{\rm BM}_{\rm template}$ from the other resonant family, i.e., the template that
maximizes the overlap ${\cal O}$ of Eq.~(\ref{overlapdefinition}).  Within our numerical precision, each member of the pair
is the other's best match; the two curves are symmetric about the diagonal $\xi_{\rm source} = \xi^{\rm BM}_{\rm template}$. 
Maximized overlaps (the tips of the peaks shown in Fig.~\ref{xi_ov_notmax}) are shown on the color scale.
}
\label{xi_xi_max}
\end{figure}

\subsection{GW power versus frequency for resonant binaries}

BBHs emit GWs anisotropically, and thus the observed response $h(t)$ depends on the direction from which they
are viewed.  We illustrate this point in Fig.~\ref{fig:expSNR}, where we plot the SNR accumulated per unit frequency
as a single pair of BBHs from each resonant family inspirals from $f = 10$
 to 300~Hz.  The top row of panels
shows $d\rho^2/df$ for the member of the $\Delta \Phi=0^\circ$ family for which $\xi = -0.5$, while the bottom row
shows the same quantity for the member of the $\Delta \Phi= \pm180^\circ$ family with the {\it same} value of $\xi$.
Each column corresponds to a different line of sight $\hat{\mathbf{n}}$, inclined with respect to the orbital angular
momentum $\mathbf{L}$ by an angle $\iota=\arccos(\hat{\mathbf{L}} \cdot \hat{\mathbf{n}})$ at the reference
frequency $f_{\rm ref}$.  These plots are normalized assuming a luminosity distance $D = 1$~Mpc, but this
normalization cancels out in the overlap ${\cal O}$ as defined in Eq.~(\ref{overlapdefinition}) that is the focus of this paper.

The first thing to note in this figure is that the GW signal depends strongly on the inclination $\iota$.
The large oscillations in $d\rho^2/df$, particularly prominent in the top row for large inclinations, result from
precession of the orbital plane about the total angular momentum.  The separation of the orbital and precessional
timescales ($t_{\rm orb} \ll t_{\rm pre}$) implies that the peaks and troughs in $d\rho^2/df$ correspond to orbital
frequencies where $\hat{\mathbf L}$ points closest to and farthest away from the line of sight (i.e., where
$|\hat{\mathbf L} \cdot \hat{\mathbf n}| = |\cos \iota|$ has local maxima and minima).  The second thing to note in
Fig.~\ref{fig:expSNR} is that these precessional oscillations are much more pronounced for BBHs in the $\Delta
\Phi=0^\circ$ family than those in the $\Delta \Phi=180^\circ$ family.  This behavior follows from the fact, explained in
Sec.~\ref{SS:diffdyn} and illustrated in Fig.~\ref{Jn_iota}, that the total spin and hence the orbital angular momentum
are more misaligned with the total angular momentum for BBHs in the $\Delta \Phi=0^\circ$ family than those in the
$\Delta \Phi=180^\circ$ family.  The very modest misalignment between $\hat{\mathbf L}$ and $\hat{\mathbf J}$
(only $\sim5^\circ$) for the $\Delta \Phi=180^\circ$ BBHs in Fig.~\ref{fig:expSNR} implies that there is little precessional modulation for any inclination $\iota$.

Computational limitations make an exploration of all possible viewing angles prohibitive, so we instead
choose $\iota$ such that the line of sight $\hat{\mathbf{n}}$ is aligned with the {\em total} angular momentum
$\mathbf{J}$.  This is a conservative choice, in that $\hat{\mathbf L}\cdot \hat{\mathbf n} = \cos \iota$ remains nearly
constant throughout the inspiral. Therefore the precessional modulations (and thus the differences between the
waveforms of the two resonant families) are minimized.

\subsection{Comparing binaries from different resonant families}
\label{SS:CompFam}

The different $d\rho^2/df$ curves shown in the top and bottom rows in Fig.~\ref{fig:expSNR} suggest that the two
resonant families may produce observationally distinguishable GW signals, even under the conservative assumption 
that both are viewed along $\hat{\mathbf n} = \hat{\mathbf J}$.  To test this hypothesis quantitatively, we compare the
GWs from each member of one family with all members of the other, using the overlap defined in
Eq.~(\ref{overlapdefinition}) as a measure of their distinguishability.
Our only parameter along each resonant family is the projected effective spin $\xi$, which describes the orientation
of {\it both} spins when locked into a coplanar configuration. For simplicity, in this case study all other binary
parameters (including masses, sky location and precession phase) are held fixed.

The largest overlap between a signal and all members of some model space provides a simple way to estimate
whether that signal with some SNR is compatible with the model space
\cite{2013PhRvD..87b4004C,2008PhRvD..78l4020L,2009PhRvD..79l4033R}.
Roughly speaking, if the largest value of the overlap ${\cal O}$ is sufficiently small [Eq.~(\ref{eq:distinguish})] the
signal can be differentiated from all members of the model family.  For the first few GW detections, we anticipate
$\rho \simeq 10$; larger amplitudes will occur as well, albeit with low probability [the cumulative distribution $P(\rho > 
\rho_0) \propto 1/\rho_0^3$].  We therefore require ${\cal O} < 0.99$ as a criterion to reliably distinguish the two
model families.

Figure~\ref{xi_ov_notmax}  shows the behavior of the overlap for five different sources in each family. Each curve
has a unique maximum; in other words, there is a one-to-one correspondence between each source $\tilde h_0(f)$ in
one of the resonant families and the template $\tilde h(f)$ in the other family which is its best approximation.  The
nature of this pairing relation is explored in Fig.~\ref{xi_xi_max}.  For each value of $\xi_{\rm source}$ from each
resonant family, we maximize the overlap ${\cal O}$ over all members of the other family and plot the value
$\xi_{\rm template}^{\rm BM}$ for this best matching template.  Within our numerical precision, the two curves in
Fig.~\ref{xi_xi_max} (one for each source family) are symmetric about the diagonal $\xi_{\rm source} =
\xi_{\rm template}^{\rm BM}$, implying that each member of the pairings between the resonant families is each other's best
match.
As explained in Appendix \ref{ap:Symmetry}, this symmetry is only approximate, but it holds throughout our restricted
parameter space to better than a part in $500$.

Figure~\ref{xi_xi_max} also shows the values of the overlaps for each pair on a color scale.  Overlaps range in value from a
minimum 
$\sim 0.93$
for binaries with effective spins nearly perpendicular to the orbital angular momentum ($\xi \sim
0$) to a maximum of unity for fully aligned or antialigned binaries [$\xi = \pm (\chi_1+q\chi_2)/(1+q) = \pm1$ for $\chi_1
= \chi_2 = 1$] belonging to both families (lower left and upper right corners of Fig.~\ref{res_S0L}).  Within the scope of
this initial study, where only one intrinsic parameter (the projected effective spin $\xi$) is allowed to vary, our conclusions
are optimistic: except for nearly aligned or antialigned binaries, GWs from the two resonant families have small enough
overlaps that they can be distinguished from one another at SNRs $\rho \gtrsim 10$ (${\cal O} \lesssim 0.99$).

This conclusion holds despite our conservative assumption that BBHs are viewed from the least favorable direction
($\hat{\mathbf n}= \hat{\mathbf J}$) which minimizes the precession-induced modulations present in both families.
To test the robustness of our conclusions, we performed a similar analysis considering generic, but fixed, values of $\iota$. As expected,  resulting overlaps decrease quite significantly, with a minimum at $\sim0.9$. The range of $\xi$ where ${\cal O} \lesssim 0.99$ moves from $-0.8\lesssim \xi \lesssim 0.8$ (Fig.~\ref{xi_xi_max}) to $-0.9\lesssim \xi \lesssim 0.9$.

\subsection{Distinguishing BBH formation scenarios}

\begin{figure}
\includegraphics[width=\columnwidth]{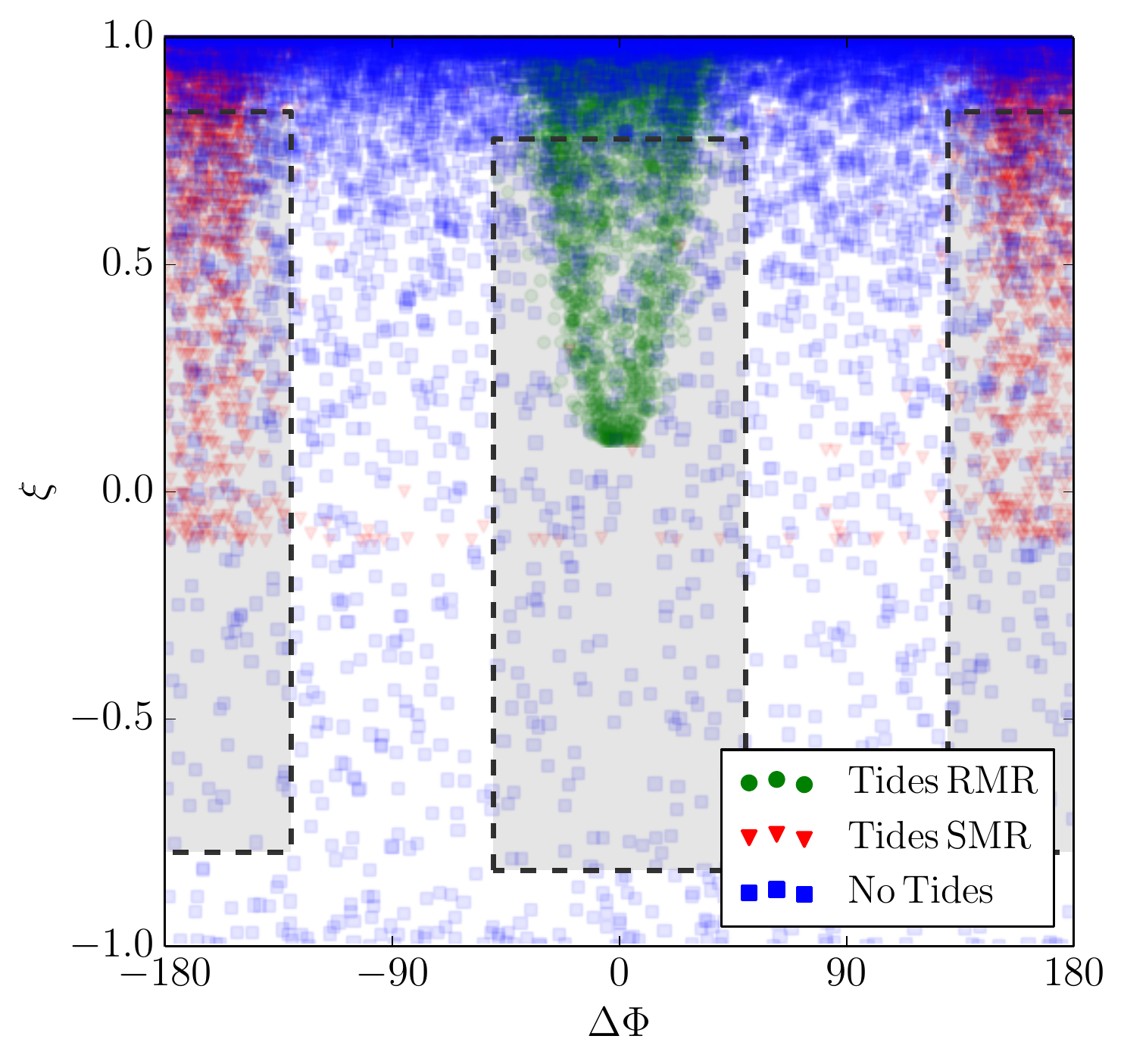}
\caption{\label{fig:apart}(color online) Distributions of $\xi$ and $\Delta\Phi$ at $f_{\rm ref}=60$ Hz in the astrophysical models we
developed in Ref.~\cite{2013PhRvD..87j4028G}.  All scenarios shown here assume isotropic supernova kicks.
Binaries for which tides align the spin of the secondary with the orbital angular momentum prior to the second
supernova are typically locked into resonances by the end of the inspiral.  When mass transfer prior to the first
supernova causes the secondary to form the more massive BH (``Tides RMR", green circles), the BBHs tend to be attracted
to the $\Delta\Phi=0^\circ$ resonances.  If this mass-ratio reversal does not occur (``Tides SMR", red triangles), binaries
will instead fall into the $\Delta\Phi=180^\circ$ resonances.  Without this tidal alignment (``No Tides", blue squares), BBHs
will show no preference for either resonant family.  BBHs inside the dashed boxes are within $\pm50^\circ$ degrees of
either $\Delta\Phi=0^\circ$ or $\Delta\Phi=180^\circ$ at $f_{\rm ref}$ and have maximum overlaps below $0.99$ with
the other family.}
\label{F:DPxidist}
\end{figure}

\begin{figure}
\includegraphics[width=\columnwidth]{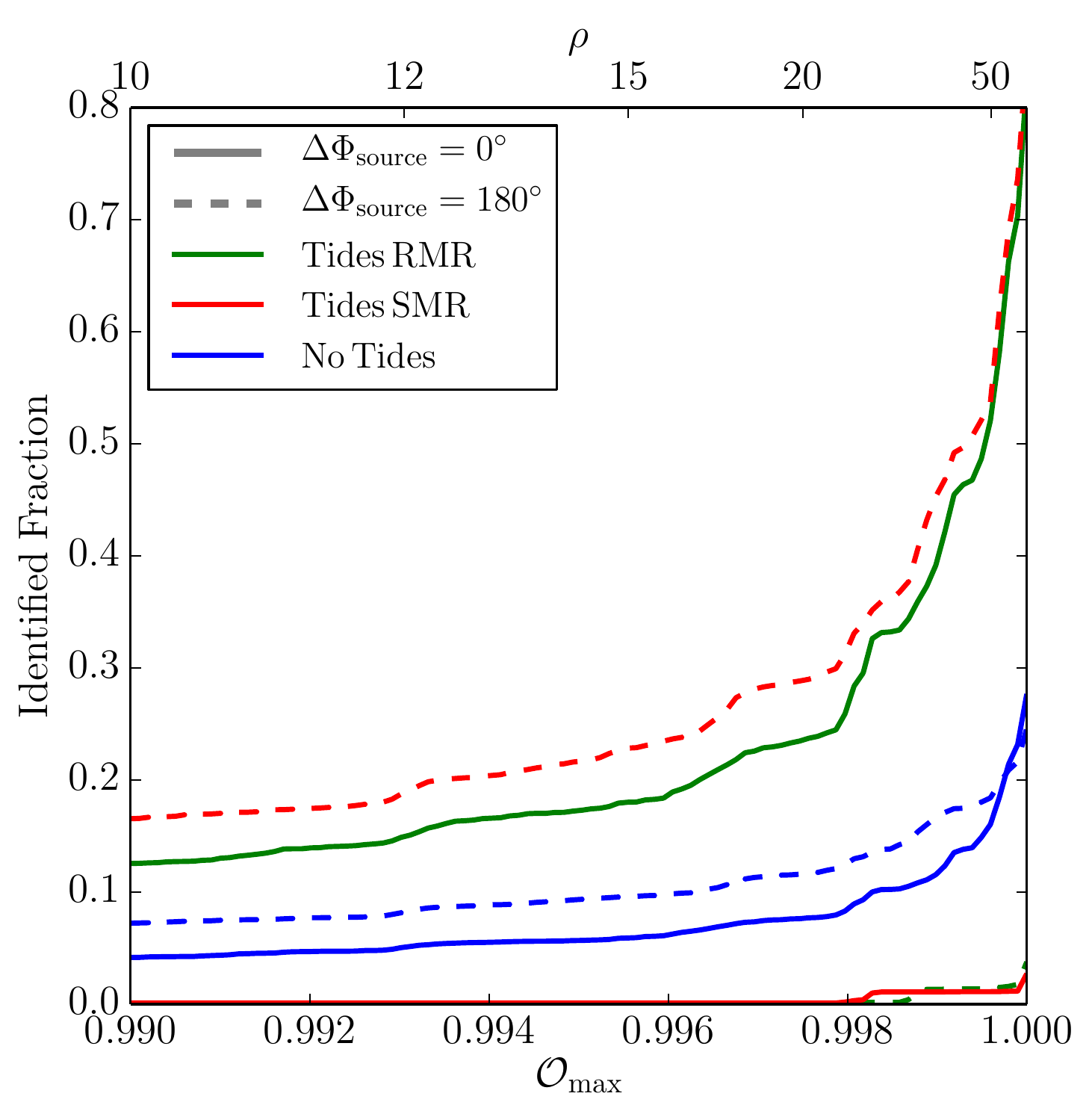}
\caption{\label{fig:distinguish}(color online) Fraction of binaries from the astrophysical distributions shown in
Fig.~\ref{F:DPxidist} that can be identified as belonging to one of the resonant families as a function of their maximum
allowed overlap ${\cal O}_{\rm max} = 1 - \rho^{-2}$ with their match in the opposite family.  As the SNR $\rho$ increases,
the range of values of $\xi_{\rm source}$ with $\cal{O} < \cal{O}_{\rm max}$ increases as seen in Fig.~\ref{xi_xi_max}.  This
range determines the heights of the dashed boxes shown in Fig.~\ref{F:DPxidist}; as the areas of these boxes increase so
too does the fraction of the points contained within them.  The solid (dashed) curves show the fraction of binaries contained
within the box centered on $\Delta\Phi = 0^\circ~(180^\circ)$.  The green, red, and blue curves correspond to the ``Tides
RMR", ``Tides SMR", and ``No Tides" scenarios respectively.  We see that virtually all identified binaries belong to the 
$\Delta\Phi = 0^\circ~(180^\circ)$ family in the Tides RMR (SMR) scenario, while a comparable fraction of identified
binaries belong to each family in the No Tides scenario.}
\end{figure}

As summarized in the introduction and discussed at great length in our previous paper \cite{2013PhRvD..87j4028G},
aspects of the astrophysics of BBH formation can profoundly influence the fraction of binaries captured into each of the
two resonant families.  If these fractions can be measured in a sample of detected GW sources, we can observationally
constrain the astrophysics of BBH formation.  We propose that the family of a resonant binary detected with SNR $\rho$
can be identified if the overlap $\cal{O}$ with its match in the other family is less than ${\cal O}_{\rm max}=1-\rho^{-2}$. 

To illustrate how these identified fractions can distinguish different BBH formation scenarios, we consider several of the
astrophysical distributions of BBH spin orientations derived in our previous paper \cite{2013PhRvD..87j4028G}.
Figure~\ref{fig:apart} shows a scatter plot of projected effective spin $\xi$ vs the angle $\Delta\Phi$ between spin
components in the orbital plane at $f_{\rm ref} = 60$~Hz for the RMR, SMR, and ``No Tides" scenarios summarized in
the introduction (see also \cite{2013PhRvD..87j4028G}).  These three distributions are readily distinguishable by eye, but
the parameters $\xi$ and $\Delta\Phi$ may not be measured precisely for sources with a modest SNR $\rho$, and we may
not be lucky enough to obtain such large samples.  We consider points shown in Fig.~\ref{fig:apart} to belong to the 
$\Delta\Phi = 0^\circ$ or $180^\circ$ resonant families if their value of $\Delta\Phi$ is within $\pm50^\circ$ of the value
of the given family.\footnote{We fix this threshold by visual inspection of the typical amplitude of librations about
resonance at $f_{\rm ref}$.}  For each of these resonant points, we determine the value of $\xi$ and calculate the
overlap $\cal{O}$ of a resonant binary with this value of $\xi$ with its match in the opposite family, as shown in
Fig.~\ref{xi_xi_max}. For a given SNR $\rho$, we count the fraction of points for each distribution and each family with
${\cal O} < {\cal O}_{\rm max} = 1-\rho^{-2}$.

The resulting six curves as a function of ${\cal O}_{\rm max}$ are shown in Fig.~\ref{fig:distinguish}.  Unfortunately, many
of the resonant binaries have such large values of $\xi$ that it is difficult to distinguish the two families (see the upper
right corner of Fig.~\ref{xi_xi_max}).  However, with enough sources one could still hope to distinguish the three
distributions.  For example, of 100 binaries detected at $\rho = 10$ (${\cal O}_{\rm max} = 0.99$), $\sim 20$ would be
expected to be found in the $\Delta\Phi = 180^\circ$ family in the SMR scenario, $\sim 15$ would be found in the
$\Delta\Phi = 0^\circ$ family in the RMR scenario, and a few would be found in {\it each} family in the ``No Tides"
scenario.  Although much work remains to be done, this example illustrates the primary result of this paper: the
astrophysics of BBH formation can be constrained by a realistic number of BBHs detected at realistic SNR.

\section{The single-spin approximation}
\label{sec:SingleSpin}

We found in Sec.~\ref{sortingresonances} that spin-orbit resonances can be parametrized by a single projected
effective spin $\xi$, then showed in Sec.~\ref{SS:CompFam} that there is a one-to-one correspondence between
binaries in the two resonant families.  This parametrization and correspondence suggest that binaries in both families
may be well approximated by binaries with a single (effective) spin.  If this approximation is valid, it may allow us to
develop semianalytic expressions for the binary orbits, accompanying GW signals, and overlaps ${\cal O}(h_o,h)$ which
will enable a far more computationally efficient exploration of the higher-dimensional parameter space for resonant
BBHs.

A good fit between nearly nonprecessing and precessing binaries occurs when their secular-phase evolution is
similar \cite{2012PhRvD..86f4020B}.  Expressions for the secular GW phase from single-spin binaries have been
derived in previous work (cf. \cite{2012PhRvD..86f4020B} and references therein).  Adopting the single-spin
approximation, we seek simple conditions to estimate when members of different resonant families will produce similar
GW signals.

If the single-spin approximation is too successful, it will raise a new question: if resonant binaries resemble binaries
with a suitably chosen single effective spin too closely, how can we tell these two classes of sources apart?  This
question will require further investigation before we can claim with confidence that semianalytic expressions can be
used to distinguish resonant families from each other and from their single-spin cousins.

\subsection{Orbital and precessional contributions to the GW phase}

The physics of single-spin binaries is summarized in \cite{1994PhRvD..49.6274A,2012PhRvD..86f4020B}.
The orbital angular momentum $\mathbf{L}$ precesses about the total angular momentum $\mathbf{J}$,
\begin{align} \label{E:simprec}
\frac{d\mathbf{L}}{dt} = \Omega_L \hat{\mathbf J}\times \mathbf{L}~,
\end{align}
with precessional frequency $\Omega_{L}$.
We define $\beta$ to be the (instantaneous) opening angle of the
precession cone,
\begin{align} \label{E:open}
\cos\beta\equiv \hat{\mathbf J} \cdot \hat{\mathbf L}\,,
\end{align}
and $\alpha$ to be the accumulated precessional phase of $\hat{L}$
about $\hat{J}$:
\begin{align}
\alpha(f)-\alpha_{\rm ref}\equiv \int_{f_{\rm ref}}^f \Omega_L(f')\, \frac{dt}{df'}\,df' \,.
\label{intalpha}
\end{align}
Here, $\alpha_{\rm ref}$ is a reference value at $f=f_{\rm ref}$, and,
in practice, $\alpha(t)$ can be computed from a knowledge of the
(time-dependent) angular momentum of the binary.

Following \cite{2012PhRvD..86f4020B,2014PhRvD..89d4021L}, we
decompose the GW signal from a precessing binary using the following
expression [Eq. (4) of \cite{2014PhRvD..89d4021L}]:
\begin{eqnarray}
h_+(t)  = \frac{2 M \eta}{D}v^2 \text{Re} \left[ z(t) e^{2i (\phi^{\rm orb}-\zeta)} \right]
\label{hplus}
\end{eqnarray}
In this expression, $\phi^{\rm orb}$ is the orbital phase and $\zeta,\,z$ are
quantities set by the orientation of $\hat{\mathbf L}$ in
the radiation frame.  The GW phase of a precessing binary
$[2(\phi^{\rm orb}-\zeta) - i\ln z)]$ can be decomposed into three parts.
First, the orbital phase ($\phi^{\rm orb}$) accumulates monotonically on the
orbital timescale.  On the precessional timescale, $\zeta$ and $\ln z$
combine to produce both periodic modulations and a \emph{secular}
increase in the overall phase, in proportion to the number of
precession cycles.  Following Brown et {\it al.}~\cite{2012PhRvD..86f4020B}, we estimate
this secular contribution by the factor $W$ such that
\begin{eqnarray}
\left< \frac{d}{dt} (-\zeta -\frac{i}{2}\ln z) \right> \simeq W d\alpha/dt,
\end{eqnarray}
where angles denote averaging over several binary orbits. 
The quantity $W$ depends on the orientation of the precession cone of $\mathbf{L}$ about $\mathbf{J}$, relative to
the line of sight; it therefore evolves on the radiation-reaction timescale. For reasons
explained above, in this work we focus on binaries seen directly along their total angular momentum
($\hat{\mathbf n}=  \hat{\mathbf J}$). Therefore, precession-induced modulations can be expected to be small. We will neglect these modulations, instead
emphasizing the secular phase $\phi^{\rm wave}$:
\begin{align}
\phi^{\rm wave}\simeq 2(\phi^{\rm orb} + W \alpha) \,.
\label{phiwave}
\end{align}
Brown et {\it al.} \cite{2012PhRvD..86f4020B} provide an exhaustive discussion of the function $W$. Since the misalignment between the line of sight and the orbital angular
momentum is always quite small (see Fig.~\ref{Jn_iota}), the line of sight never crosses the orbital plane during the inspiral.
 Whenever this condition holds (and it does, as we checked numerically by integrating the PN equations), the quantity $W$ assumes the simple form  \cite{2012PhRvD..86f4020B} 
\begin{align}
W={\rm sign}(\cos\beta)-\cos\beta \,.
\end{align}

\begin{figure*}
\includegraphics[width=0.49\textwidth]{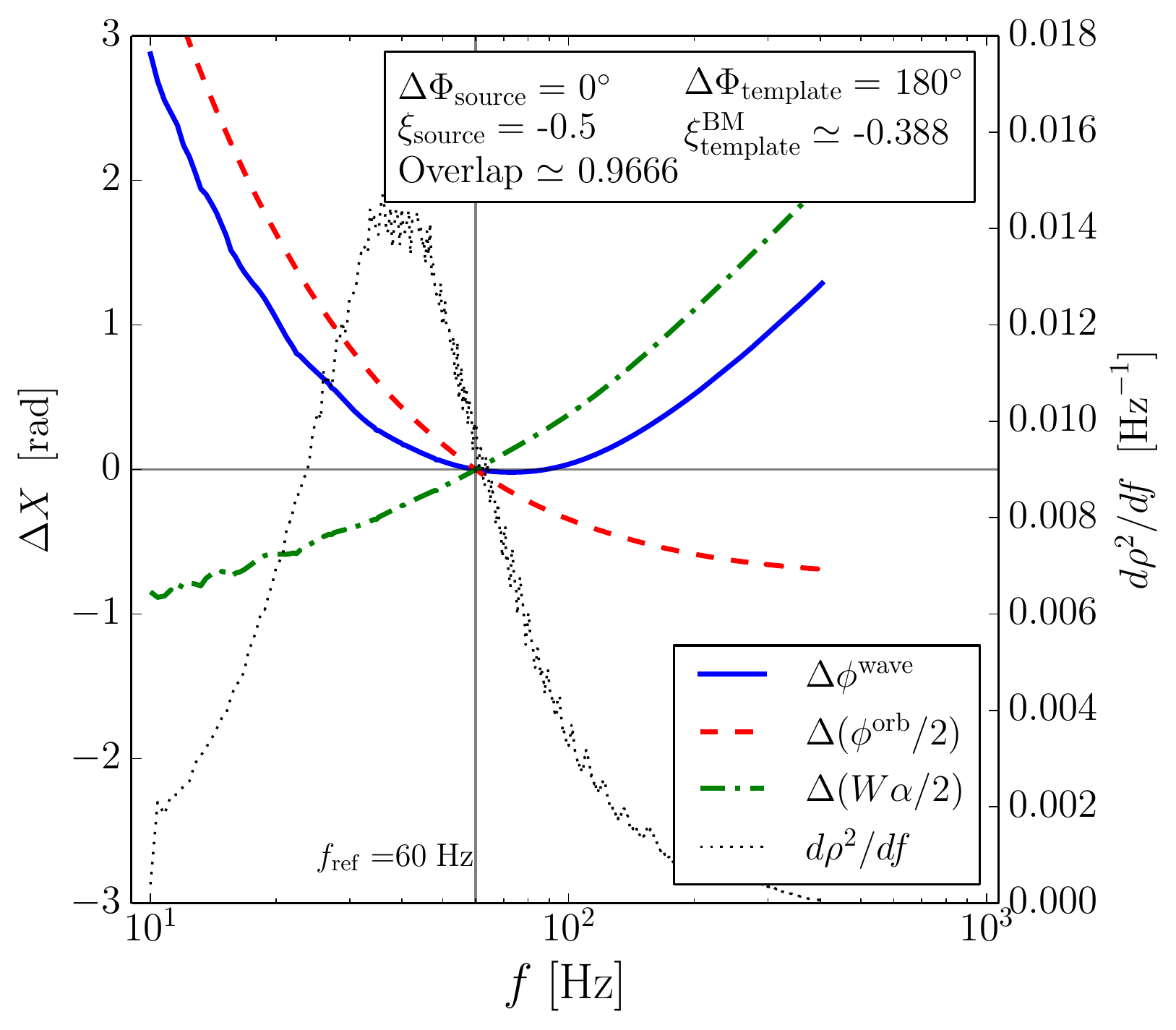}
\includegraphics[width=0.49\textwidth]{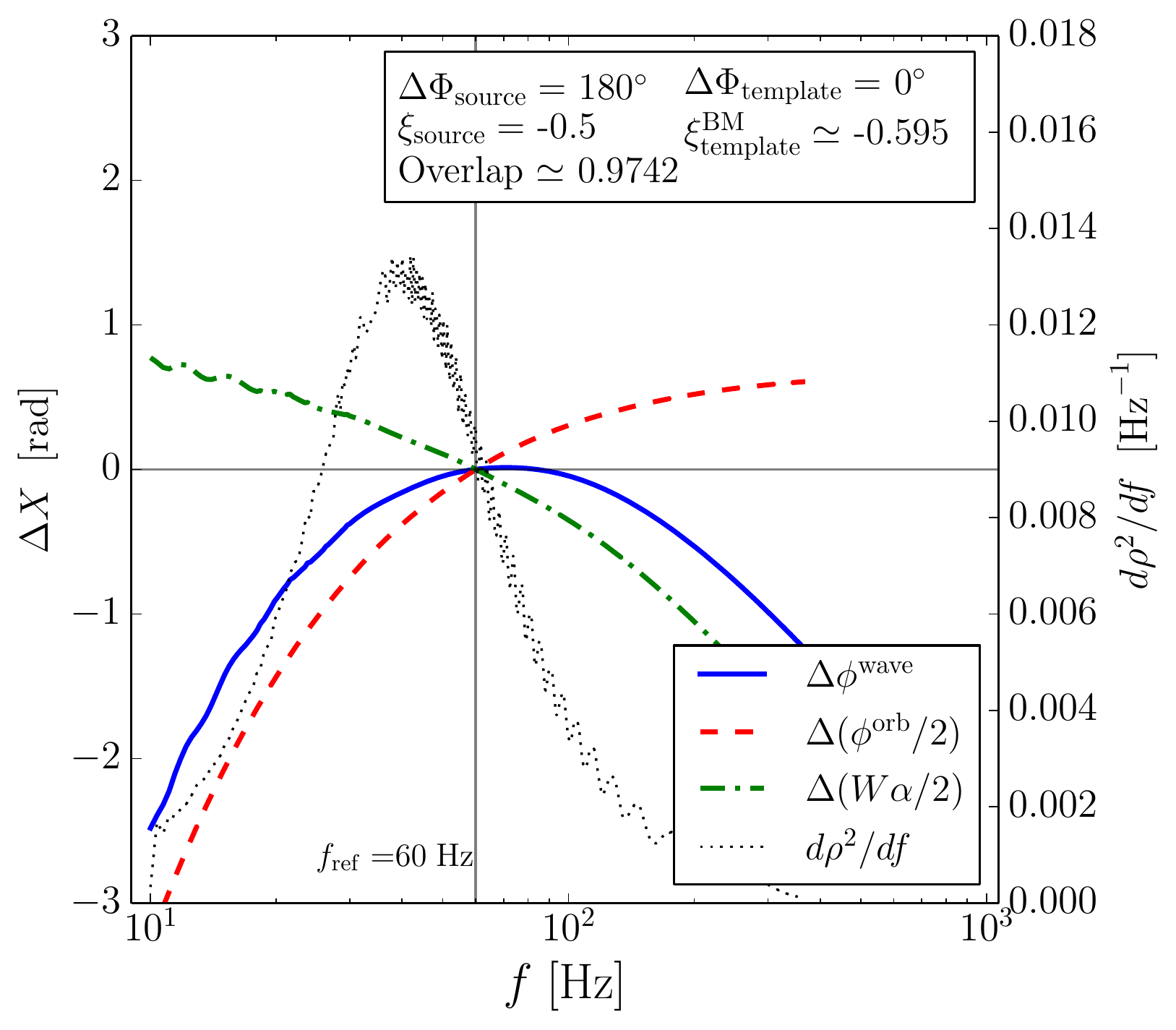}
\caption{(color online) Phasing of best matching BH binaries using the
  single-spin approximation, for a single source in the
  $\Delta\Phi=0^\circ$ family (left) and in the $\Delta\Phi=180^\circ$
  family (right).  We denote by $\Delta X$ (left axis in each panel) the difference between the
  quantity $X$ for the source and for the best matching template, as a
  function of the GW frequency $f$. The solid blue line shows the GW
  phase offset between the source and the best matching template
  computed within the single-spin approximation (\ref{phiwave}). The
  dashed red and dot-dashed green lines show the orbital and the
  precessional contributions, respectively. The dotted black lines
  show the source squared SNR per unit frequency assuming a luminosity distance $D=1~{\rm Mpc}$, as reported on the right axis in each panel.}
\label{phiwavefig}
\end{figure*}

Both the phase $\phi^{\rm orb}$ and the accumulated precession $\alpha$ are defined to be zero at the reference
frequency: the former by choosing $\phi_{\rm ref}=0$, the latter by fixing $\alpha_{\rm ref}=0$ in Eq.~(\ref{intalpha}).  It
follows that $\phi^{\rm wave}=0$ at $f_{\rm ref}$. The evolution of $\phi^{\rm wave}$ with the frequency is
computed numerically by integrating forwards in time for $f>f_{\rm fref}$, and backwards in time for $f<f_{\rm ref}$.
Figure \ref{phiwavefig} shows the evolution of the GW phase $\phi^{\rm wave}$ during the inspiral for two resonant
sources.  The solid blue curves show the difference $\Delta\phi^{\rm wave}$ between the secular GW phase
$\phi^{\rm wave}$ for one source from each family and its match in the opposite family.  The
contributions to this difference from the orbital phase $\phi^{\rm orb}$ and secular precession $W\alpha$ [the two
terms in Eq.~(\ref{phiwave})] are shown with dashed red and dot-dashed green curves, respectively.  We also show 
for comparison the squared SNR per unit frequency of the source.

\subsection{Two matchmakers for the resonant families}
\label{SS:2match}

We found in Sec.~\ref{SS:CompFam} that each resonant binary has a best fitting companion in the other
family.
In this section, we use the single-spin approximation to develop two different criteria for predicting which pair of resonant 
binaries in each family will be each other's best match.  The accuracy of these predictions will test the
validity of the single-spin approximation and our understanding of the GW phenomenology that governs high overlaps.

\subsubsection{Effective number of cycles}

Damour et {\it al.} \cite{2000PhRvD..62h4036D} define $\mathcal N(f)$
as the differential (logarithmic) contribution to the total number of
cycles
\begin{align}
\frac{1}{2\pi}\int d\phi^{\rm wave}=\int \frac{df}{f} \mathcal N(f)\,,
\label{Nwave}
\end{align}
where $\phi^{\rm wave}$ is the phase defined in Eq.~(\ref{phiwave}) for the single-spin approximation. The source's SNR can be expressed as a weighted integral over $\mathcal N(f)$,
\begin{align}
\rho^2= \int \frac{df}{f} \mathcal N(f) w(f)\,,
\end{align}
where the weighting $w(f)$ is given by
\begin{align}
w(f)&=2\pi \frac{d \rho^2}{df} \left( \frac{d\phi^{\rm wave}}{df}\right)^{-1}
\notag \\
&=8\pi \frac{|\tilde h_0 (f)|^2}{S_n(f)} \left( \frac{d\phi^{\rm wave}}{df}\right)^{-1}.
\end{align}
We can use the differential number of cycles $\mathcal{N}(f)$ and weighting $w(f)$ to define an effective number of
cycles
\begin{align}
N_{\rm eff}=\left[ \int \frac{df}{f} \mathcal N(f) w(f) \right] \left[ \int \frac{df}{f} w(f)\right]^{-1}~.
\label{effcycle}
\end{align}
Inspired by this definition of $N_{\rm eff}$, our first prediction for the match to a source in one family with
$\mathcal{N}_s(f)$ and weighting $w_s(f)$ is the template in the other family with $\mathcal{N}_t(f)$
that minimizes
\begin{align} \label{E:cycdiff}
|\Delta{\mathcal N}| =  \left| \int \frac{df}{f} [{\mathcal N}_s(f) - {\mathcal N}_t(f)] w_s(f) \right| \left[ \int \frac{df}{f} w_s(f)\right]^{-1}~.
\end{align}

\subsubsection{Linearized phasing}

In the single-spin approximation, the GW signal can be decomposed into a time-varying amplitude and GW phase
(\ref{hplus}).  The dephasing ${\cal Q}$ between a source and template whose secular GW phase (\ref{phiwave}) differs by
an amount $\Delta \phi^{\rm wave}$ and whose coalescence phase and arrival time differ by $\phi_c$ and $t_c$ is given
by
\begin{align} \label{E:dephase}
{\cal Q}(\phi_c,t_c)=\text{Re} \langle  e^{ i\left(\Delta \phi^{\rm wave} - 2\pi f t_c - \phi_c\right)}\rangle~,
\end{align}
where angle brackets denote the SNR-weighted average
\begin{align}
\langle X \rangle =\left[\int df \frac{d \rho^2}{df} X \right]\left[ \int df \frac{d  \rho^2}{df} \right]^{-1}~.
\end{align}
If the source and template have the same amplitude [or equivalently the same squared SNR per unit frequency
$d\rho^2/df$ (\ref{drho2df})], their overlap will be given by
\begin{eqnarray}
{\cal O} = \text{max}_{t_c,\phi_c} {\cal Q}(t_c,\phi_c)~.
\end{eqnarray}
If this overlap is large, as is the case between the best matching pairs of resonant binaries, we can Taylor expand the
argument of the exponential in Eq.~(\ref{E:dephase}) and keep only the lowest-order real terms:
\begin{align} \label{E:OverTay}
{\cal O} \simeq  \max_{\phi_c t_c} \left\langle
\left[ 1- \frac{1}{2}\left( \Delta\phi^{\rm wave}- 2\pi f t_c -\phi_c \right)^2\right] \right\rangle~.
\end{align} 
This expansion allows us to analytically determine the values of $\phi_c$ and $t_c$ that maximize ${\cal O}$ by setting
$\partial {\cal Q}/\partial\phi_c = \partial {\cal Q}/\partial t_c=0$.  Doing so and substituting the resulting values back into
Eq.~(\ref{E:OverTay}), we find
\begin{align}
{\cal O}  &\simeq 1- \frac{1}{2}\phi^{2}_{\rm rms} \label{phirmsdef}
\end{align}
where\footnote{Note that in general $\langle X \rangle^2 \neq \langle X^2 \rangle$.}
\begin{align}
\phi^{2}_{\rm rms} &= \langle \Delta{\phi^{\rm wave\,}}^2\rangle -\langle \Delta{\phi^{\rm wave\,}}\rangle^2 
\notag\\
&- \frac{
\left( \langle f   \Delta{\phi^{\rm wave\,}}\rangle -\langle f \rangle\langle \Delta{\phi^{\rm wave\,}}\rangle \right)^2
}{
\langle f^2\rangle -\langle f\rangle^2
}\,.
\label{phirms}
\end{align}

\begin{figure*}
\includegraphics[width=0.48\textwidth]{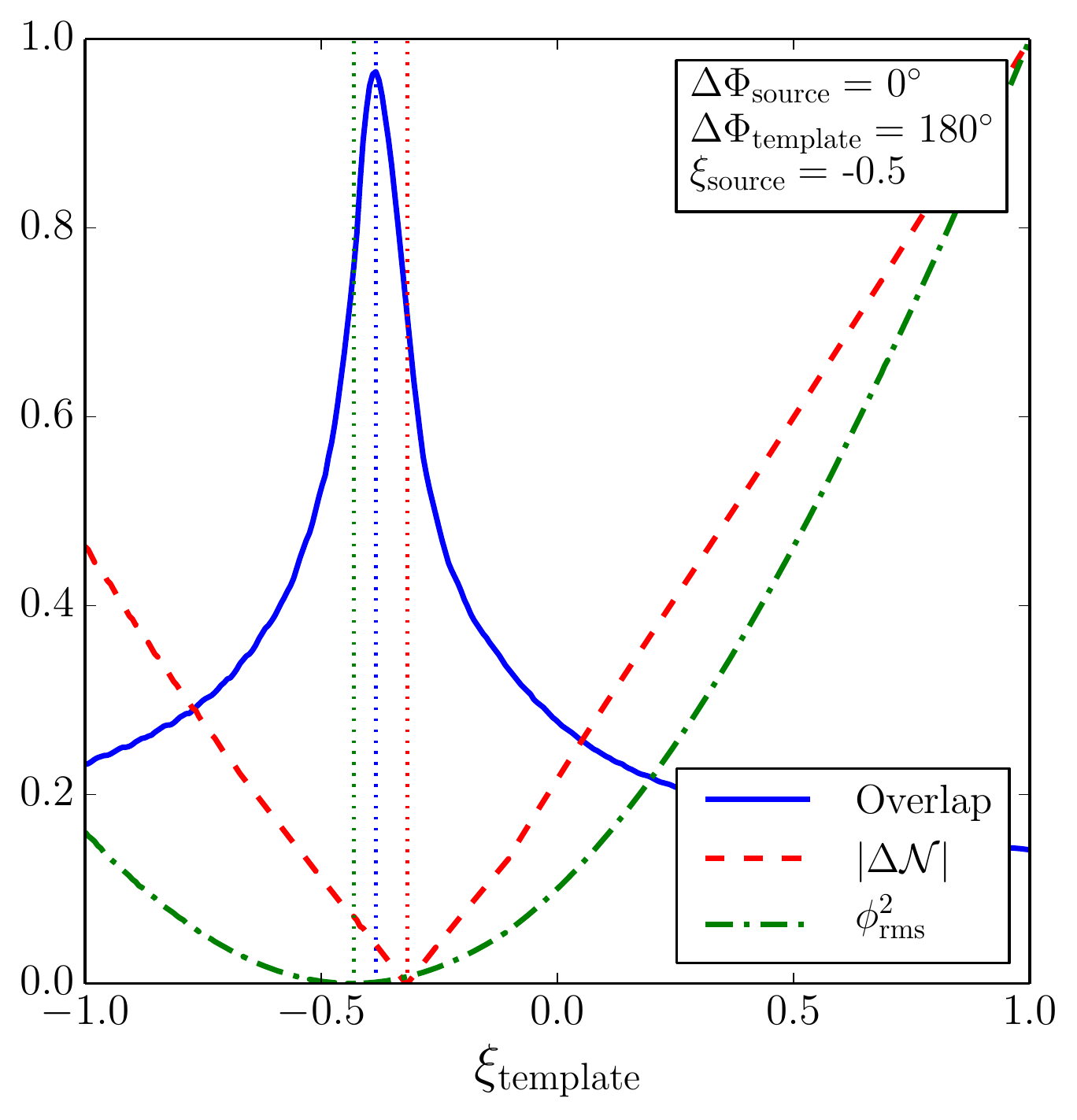}
\includegraphics[width=0.48\textwidth]{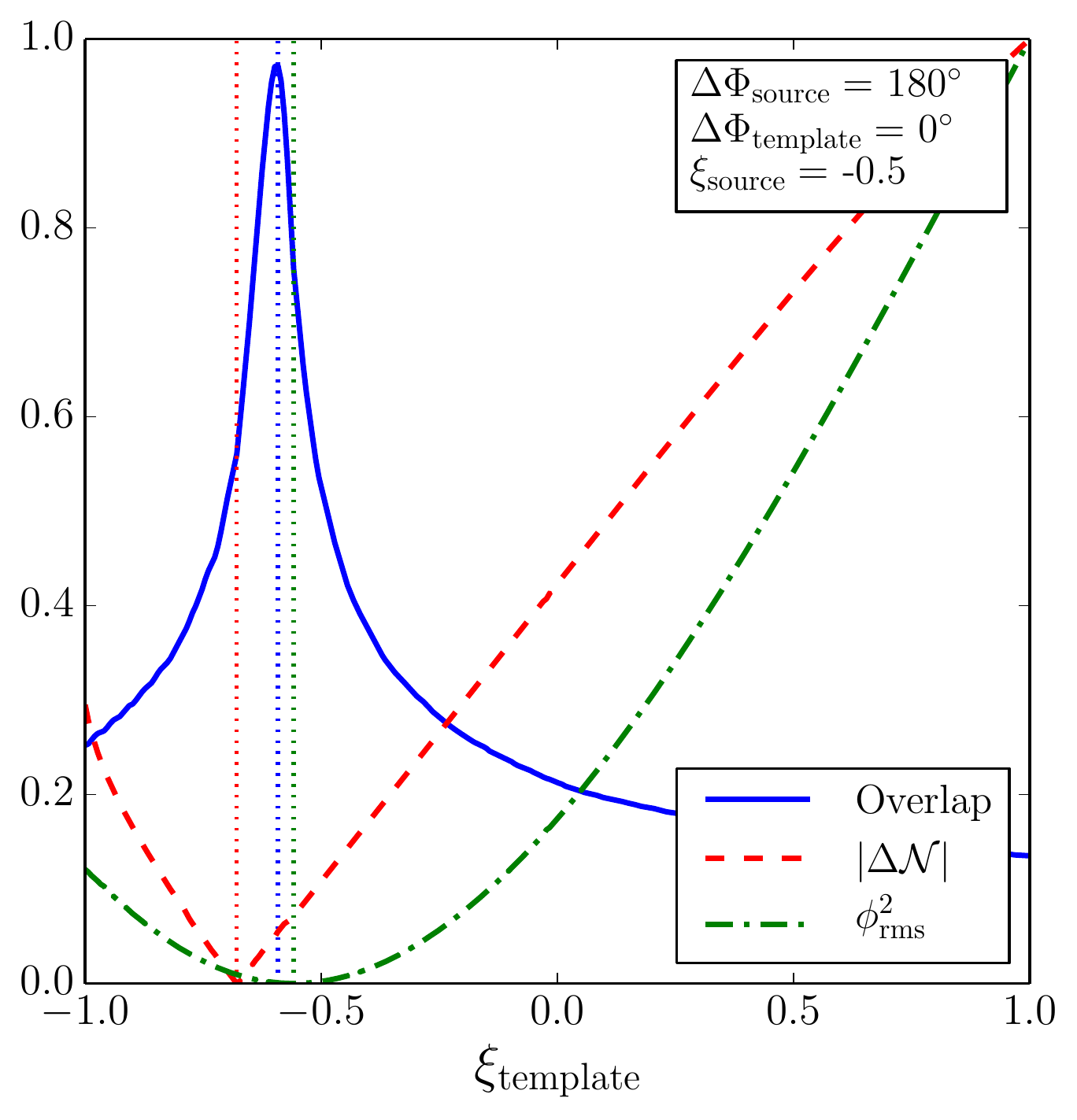}
\caption{(color online) Predictions for the best matching binaries using the single-spin approximation. Solid blue lines show the
overlap between a fixed source (with $\Delta\Phi=0^\circ$ on the left and $\Delta\Phi=180^\circ$ on the right) and
different templates from the other family. The dashed red curves show the offset $|\Delta\mathcal N|$ of the effective
number of cycles (\ref{E:cycdiff}) between the fixed source and each template. The dash-dotted green curves show the
approximate dephasing $\phi^{2}_{\rm rms}$ (\ref{phirms}); both $|\Delta \mathcal{N}|$ and
$\phi_{\rm rms}^2$ are shown in arbitrary units. Vertical dotted lines show the best matching template and the 
predictions for this template using the criteria $\min |\Delta\mathcal N |$ and $\min \phi^{2}_{\rm rms}$
described in Sec.~\ref{SS:2match}.}
\label{predictors}
\end{figure*}

\begin{figure}
\includegraphics[width=\columnwidth]{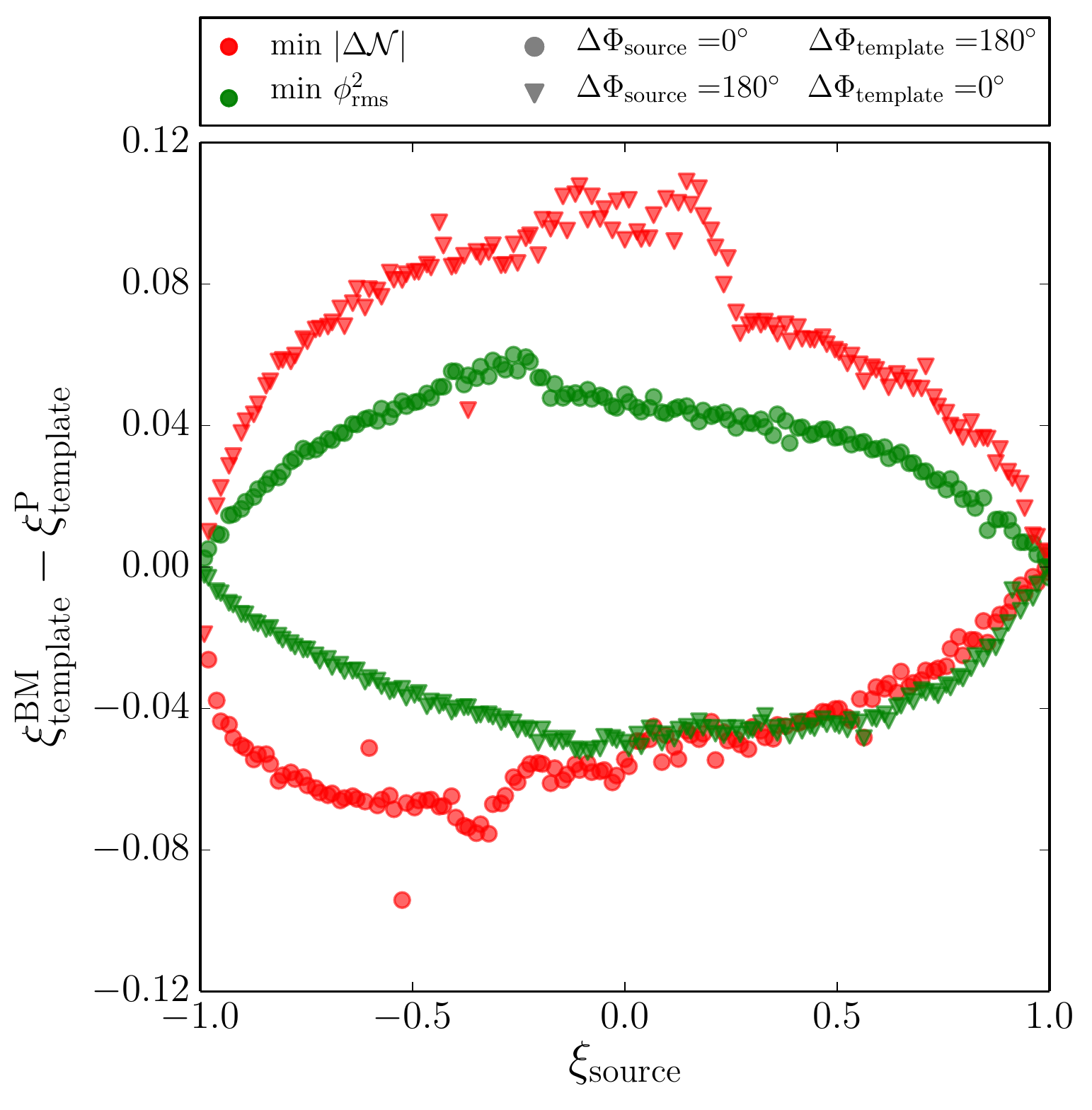}
\caption{(color online) Performance of the two predictors $\min |\Delta \mathcal N|$ and $\min \phi^{2}_{\rm rms}$. For various sources
  in both families, we show the difference between the highest-overlap template $\xi^{\rm BM}_{\rm template}$ and the
  predicted value $\xi^{\rm P}_{\rm template}$. 
}
\label{performance}
\end{figure}

Our second prediction for the template that maximizes the overlap with a source in the opposite family is therefore the
template that minimizes $\phi^{2}_{\rm rms}$.  Figure~\ref{predictors} shows the difference in the number of effective
cycles $|\Delta\mathcal N|$ (\ref{E:cycdiff}) and the mean-squared dephasing $\phi^{2}_{\rm rms}$ (\ref{phirms}) as
functions of $\xi_{\rm template}$ for $\xi_{\rm source} = -0.5$ from both families.  The templates that minimize these
two quantities are our two predictions $\xi^{\rm P}_{\rm template}$ for the best match $\xi^{\rm BM}_{\rm template}$ to
$\xi_{\rm source}$.  Figure~\ref{performance} shows the errors of these two predictions as a function of $\xi_{\rm source}$;
we see that $|\xi^{\rm BM}_{\rm template} - \xi^{\rm P}_{\rm template}| \lesssim 0.05$ for most of the parameter space,
suggesting that the single-spin approximation may be accurate enough to assess the distinguishability of the two
resonant families.

\subsection{Can double-spin resonances be distinguished from single-spin binaries?}

The success of the single-spin approximation in the previous two subsections suggests that the dynamics of resonant
binaries and the resulting GW emission do indeed resemble those of binaries with a single effective spin, perhaps related
to the spin $\mathbf{S}_0$ of Eq.~(\ref{E:S_0}) introduced to parametrize the spin-orbit resonances.  The orbital
angular momentum $\mathbf L$ in resonant binaries undergoes simple precession about $\mathbf J$ as in
Eq.~(\ref{E:simprec}) for systems with a single spin, so it is reasonable to expect that the dynamics and GW emission
might be qualitatively similar.  However, large effective spins can only be realized in comparable-mass binaries if
{\it both} BBHs have large spins.  This suggests that at least those resonant binaries with the largest effective spins
should be distinguishable from single-spin systems.

Further study is required to test the validity of this conjecture.  Double-spin systems have (small!) differences in dynamics
and radiation content (i.e., higher harmonics sourced directly by the spin) compared to truly single-spin binaries.    While we
anticipate that both features can be used to distinguish single-spin binaries from those in spin-orbit resonances, estimation of
the threshold at which these two families can be systematically distinguished is left to future work. 

\subsection{Searches in higher-dimensional parameter space}

Our analysis in this paper has been restricted to GWs emitted by binaries with comparable masses and maximal
spin magnitudes viewed from a single direction ($\hat{\mathbf n} = \hat{\mathbf J}$). When viewed from this direction,
differences in the GW signal can be traced to differences in the secular GW phase, a combination of orbital and
precession effects.  For \emph{fixed} masses and spin magnitudes, the two resonant families have different secular
GW phases for all $\xi$, ensuring that the two signals can be distinguished.

True parameter estimation requires a larger model family, including allowing the masses and spin magnitudes of
resonant binaries to vary.  We expect the maximum overlap between a source in one family and templates in the opposite
family to increase in this expanded model space.  If there are enough parameters to produce fully generic secular-phase
evolution, as suggested by \cite{2012PhRvD..86f4020B}, the overlap will approach unity provided the viewing direction
remains restricted to $\hat{\mathbf n} = \hat{\mathbf J}$.

Fortunately, most binaries will not be viewed from this privileged direction.  The GW signal in a generic direction
$\hat{\mathbf n} \neq \hat{\mathbf J}$ includes distinctive modulations that depend on the rate $\Omega_L$
[Eq.~(\ref{E:simprec})] at which $\mathbf L$ precesses about $\mathbf J$ and the opening angle $\beta$ [Eq.~(\ref{E:open})] of this
precession cone.  These quantities depend on the evolution of the BBH spin directions throughout the inspiral, and, as
we have shown in Sec.~\ref{SS:diffdyn}, two resonant families have qualitatively distinct dynamics.  We speculate the
resulting differences in the precessional modulations of the GW signal will break the degeneracies that will no doubt
exist in higher-dimensional model spaces.  
 
\section{Discussion} \label{S:disc}

BBH formation remains shrouded in mystery.  Such systems are predicted to be very rare; none have been observed to
date, which is not surprising, given their minimal electromagnetic signature.  BBH mergers are copious sources of GWs,
however, so they should be a prominent signal for GW detectors, in contrast to electromagnetic telescopes.  GW
detectors can, in principle, measure all of the intrinsic parameters associated with a binary if that binary is detected with
a sufficient SNR $\rho$.  Our previous paper \cite{2013PhRvD..87j4028G} established a surprisingly tight connection
between BBH spin orientations and BBH formation: binaries with an efficient tidal alignment that undergo a mass-ratio
reversal will preferentially be found in the $\Delta\Phi = 0^\circ$ family of resonances, those that fail to undergo such a 
reversal will preferentially be found in the $\Delta\Phi = \pm180^\circ$ family of resonances, and those without an efficient
tidal alignment are equally likely to be found in either resonant family.  A measurement of the fraction of BBHs in each
resonant family could therefore be used to distinguish between different astrophysical scenarios of BBH formation.  This
paper is the first attempt to assess the feasibility of such a proposed measurement.

The qualitatively distinct spin orientations in the two families lead to quantitative differences in the amount of orbital-plane precession.  The greater misalignment between the orbital angular momentum $\mathbf L$ and the total angular
momentum $\mathbf J$ in the $\Delta\Phi = 0^\circ$ family implies greater precessional modulation of the resulting
waveforms, even under the conservative assumption that binaries are viewed from a direction $\hat{\mathbf{n}} =
\hat{\mathbf{J}}$ where precessional modulation is minimized.  Precession-induced differences between the waveforms
generated by binaries in the two resonant families lead to a maximum overlap ${\cal O}_{\rm max}(\xi_{\rm source})
< 1$ between a source with projected effective spin $\xi_{\rm source}$ in one family and the best matching template
$\xi_{\rm template}^{\rm BM}$ from the other family.  The slow variation of ${\cal O}_{\rm max}(\xi_{\rm source})$ implies
that this matching is symmetric to better than a part in $10^3$: the binary from the first family with $\xi_{\rm source}$ is
also very nearly the template that provides the best match when the binary from the second family with
$\xi_{\rm template}^{\rm BM}$ is serving as the source.  The resonant family of a binary with $\xi_{\rm source}$ can be
identified when ${\cal O}_{\rm max}(\xi_{\rm source}) < 1- \rho^{-2}$; this condition holds for much of our one-parameter
space $\xi_{\rm source} \in [-1, +1]$ for $\rho \gtrsim 10$, a typical SNR expected for the first GW detections.\footnote{
After our study was completed, the authors became aware of a work by Vitale et {\it al.}~\cite{2014arXiv1403.0129V} that performs detailed parameter estimation on selected generic double-spin binaries.  
Unfortunately, these authors did not select resonant configurations for their detailed investigation (even if their injected configurations are coplanar at $f=100~{\rm Hz}$).
}

Different astrophysical BBH formation scenarios can be distinguished if they predict that measurably different fractions
of binaries reside in the portions of parameter space that can be identified by the criterion above as belonging to each
of the resonant families.  This is indeed the case for the three scenarios described in our previous paper
\cite{2013PhRvD..87j4028G}; if 100 binaries are detected with $\rho \gtrsim 10$, $\sim15$ should be found in the
$\Delta\Phi = 0^\circ$ family in the reverse-mass-ratio scenario, $\sim20$ should be found in the $\Delta\Phi =
\pm180^\circ$ family in the standard-mass-ratio scenario, and $\sim5$ should be found in {\it each} family if the tidal
alignment is inefficient.  These three scenarios and the resulting distributions of BBH spin orientations were constructed
long before we calculated our first overlap, and thus are in no way optimized to maximize the number of binaries in the
identifiable portion of parameter space.  
Finally, except for contrived scenarios, BBHs should  be detected frequently
\cite{2010CQGra..27q3001A,2012ApJ...759...52D}, with a rate of events at SNR $>\rho$ roughly proportional to $\simeq
O(1-1000)\unit{yr}^{-1}(10/\rho)^3$ at the design sensitivity.   Extrapolating from our results, only for pessimistic scenarios do
we expect to have too few and too faint events to distinguish between the RMR and SMR scenarios. 

Our claim that GW detectors can be used to constrain BBH formation scenarios must remain provisional until more
realistic higher-dimensional model parameter spaces are considered.\footnote{A recent paper \cite{2014CQGra..31j5017G}
has some overlap with our own and also argues that the two gravitational-wave signals can be distinguished.}  Our
demonstration that the single-spin approximation describes resonant binaries with reasonable accuracy may facilitate such
a higher-dimensional analysis, but this remains a subject for future work.  
Our current study offers the tantalizing promise that Advanced LIGO/Virgo may not only discover GWs and test general
relativity in the strong-field regime, but also may revolutionize our understanding of astrophysical BBH formation.

\section*{{ACKNOWLEDGMENTS}}
We thank  M. P\"urrer and the referee for helpful feedback.
D.G. is supported by the United Kingdom Science and Technology Facility Council and
the Isaac Newton Studentship of the University of Cambridge. E.B. is
supported by the NSF CAREER Grant No. PHY-1055103.  R.O.S. is supported by the
NSF Grants No. PHY-0970074 and No. PHY-1307429.    U.S. is supported by the FP7-PEOPLE-2011-CIG Grant No. 293412 ``CBHEO," the FP7-PEOPLE-2011-IRSES Grant No. 295189 ``NRHEP," the STFC GR Roller Grant No. ST/L000636/1, the Cosmos system, part of DiRAC, funded by STFC and BIS under Grants No. ST/K00333X/1 and No. ST/J005673/1, the NSF XSEDE Grant No. PHY-090003, and CESGA-ICTS Grant No. 249. Figures have been generated
using the \textsc{Python}-based \textsc{matplotlib} package
\citep{2007CSE.....9...90H}.

\appendix
\section{IMPLEMENTATION OF BINARY EVOLUTION AND GW EMISSION}
\label{app_coord}

We evolve quasicircular inspiral orbits and generate the associated PN GW inspiral signal using the {\sc lalsuite}
SpinTaylorT4 code, developed by the LIGO Collaboration \cite{lal} based on prior work
\cite{2003PhRvD..67j4025B,2009PhRvD..80h4043B}.  We adopt precisely the same signal model used in
\cite{2014PhRvD..89f4048O}.  The orbital-phase and frequency evolution include nonspinning corrections to 3.5PN
order, spin terms to 2.5PN order, and precession to 2PN order.  The outgoing radiation includes harmonics up to 1.5PN
order.  Memory terms are omitted.
Pairs of GW signals are compared using the zero-detuned high-power noise curve $S_n(f)$  foreseen for Advanced
LIGO \cite{2010CQGra..27h4006H,LIGO-aLIGODesign-Sensitivity,2013arXiv1304.0670L}, with a lower cutoff at
$f_{\rm min}=10$ Hz.
FFTs are computed with a default sampling rate: $\Delta T=1/4096$~s.

Quasicircular spinning BBHs and their associated GW emission are described by two sets of parameters.  {\it Intrinsic}
parameters depend on the physical properties of the source, while {\it extrinsic} parameters depend on the location and
orientation of the GW detector.  Each of these parameters must be specified at some point during the evolution of the
binary, i.e. at some reference GW frequency $f_{\rm ref}$.  As already pointed out in
\cite{2013PhRvD..87b4004C,2014PhRvD..89f4048O}, waveforms that have similar phasing at frequencies where the
detector is most sensitive will appear much more similar to each other than waveforms whose phase is matched outside
the region of peak sensitivity.  We therefore specify all binary parameters (and, in particular the spin orientation; see
Sec.~\ref{sortingresonances}) at $f_{\rm ref}=60$ Hz which is near the peak of the squared SNR per unit frequency
$d\rho^2/df$ for most of our sources.

Intrinsic parameters include the component masses $m_1$ and $m_2$ and quantities derived from them such as the
total mass $M = m_1 + m_2$, the mass ratio $q = m_2/m_1 \leq 1$, the symmetric mass ratio $\eta=m_1m_2/M^2 =
q/(1+q)^2$, and the chirp mass ${\cal M}=\eta^{3/5}M$.
The BH spins $\mathbf S_1$ and $\mathbf S_2$ are also intrinsic parameters, with magnitudes given by the usual
dimensionless spins $\chi_i = S_i/m_i^2$ with $0\leq \chi_i\leq 1$ and orientations described below.

Extrinsic parameters include two angles to define the source position in the detector's sky and another two angles to
specify the orientation of the angular momentum $\mathbf L$.  In the radiation frame (i.e., relative to the emission
direction $\hat{\mathbf n}$), these two angles are the inclination $\cos \iota= \hat{\mathbf L} \cdot \hat{\mathbf n}$
and a polarization angle $\psi$ describing the direction of $\hat{\mathbf L}$ in the plane of the sky (i.e., perpendicular to
$\hat{\mathbf n}$).  The luminosity distance $d$ between the binary and the observer is also an extrinsic parameter, but
it only sets the overall normalization of the SNR $\rho$ and thus cancels in calculations of the normalized overlap
${\cal O}$ in Eq.~(\ref{overlapdefinition}).
An orbital phase $\phi_{\rm ref}$ at $f_{\rm ref}=60\unit{Hz}$ is also required to generate waveforms, but it similarly
cancels in calculations of the overlap that are maximized over the orbital phase.\footnote{The orbital phase can be
specified at any point.  Though the coalescence phase appears in our definition of the overlap, maximization over the coalescence phase implies maximization over $\phi_{\rm ref}$.}

When specifying initial conditions at $f_{\rm ref}=60$~Hz, we work in the radiation frame where the line of sight
$\hat{\mathbf{n}}$ lies along the $z$ axis and the orbital angular momentum $\mathbf{L}$ lies in the $xz$-plane. Our geometry is summarized in Fig.~\ref{radframe}.

By construction, the angular momentum orientation is fully defined by the angle $\iota$:
\begin{align}
{\mathbf L}&=\frac{\eta M^2}{(\pi f M)^{1/3}}\left[\sin\iota \mathbf{e_x} + \cos\iota \mathbf{e_z}\right]\,.
\label{Llal}
\end{align}
The spin directions $\hat{\mathbf{S}}_i$ in this frame are each described by two angles ($\theta_i, \Phi_i$), where $\theta_i= \arccos (\hat{\mathbf{S}}_i \cdot \hat{\mathbf{L}})$, and $\Phi_i$ is defined to be the angle between the projections of $\hat{\mathbf{S}}_i$ and $\mathbf{e_x}$ on the orbital plane:
\begin{align}
{\mathbf S}_i =m_i^2 \chi_i&\left[ (\cos\iota\sin\theta_i\cos\Phi_i+\sin\iota\cos\theta_i) \mathbf{e_x} + \right.
\notag
\\
&+ (\sin\theta_i\sin\Phi_i) \mathbf{e_y} 
\notag
\\&\left. +(\cos\iota\cos\theta_i-\sin\iota\sin\theta_i\cos\Phi_i) \mathbf{e_z}\right]~.
\end{align}
Coplanar configurations correspond to 
\begin{align}
\Delta\Phi\equiv\Phi_2-\Phi_1=0^\circ , \pm 180^\circ\,.
\end{align}
The vectors ${\mathbf L}$, $\hat{\mathbf S}_1$, $\hat{\mathbf S}_2$ evolve during the inspiral;  binaries are identified
by the values of these parameters at the reference frequency $f_{\rm ref}=60$~Hz.
We fix the detector orientation by setting $\psi=0$, so that the projection of $\hat{\mathbf L}$ in the plane of the sky at
$f_{\rm ref}=60$~Hz is parallel to the detector's ``$x$'' arm.

To isolate the differences in the waveforms from the two families of resonant binaries, we fix all the binary parameters
but the \emph{relative} spin directions. As stated in the main text, we focus on BBHs with $M = 13.5 M_\odot$, $q = 0.8$ and maximal spins ($\chi_1=\chi_2=1$). Furthermore, we fix the line of sight to be along the total angular momentum, i.e.
$\hat{\mathbf n}= \hat{\mathbf J}$ at $f_{\rm ref}$. This choice was made for two reasons. The first is that it allows us to
separate resonant effects from purely geometrical effects due to the direction of observation. The second (and perhaps
most important in the present context) is that this particular configuration minimizes precessional effects that distinguish
the two resonant families, and therefore, from a GW data analysis point of view, it yields conservative predictions on the
resolvability of resonant binaries.
For coplanar configurations ($\sin \Delta\Phi=0$), the choice
$\hat{\mathbf n}= \hat{\mathbf J}$ corresponds to
\begin{align}
\cos\Phi_1 &= - \text{sgn} \left( |\mathbf{S_1}|  \sin \theta_1 + |\mathbf{S_2}|  \sin \theta_2 \cos\Delta\Phi  \right)\,, 
\label{Jalongn_phi1}
\\
\cos\iota &= \frac{ |\mathbf{L}| +  |\mathbf{S_1}| \cos\theta_1 +   |\mathbf{S_2}| \cos\theta_2 }{|\mathbf{J}|}\,.
\label{Jalongn}
\end{align}
This choice leaves only $\theta_1$, $\theta_2$ and $\Delta\Phi$ as freely specifiable parameters.

\section{APPROXIMATE SYMMETRY OF MAXIMIZED OVERLAPS}
\label{ap:Symmetry}

Let us consider two different one-parameter families of waveforms $I_1$ and $I_2$, where the waveforms
$\tilde{h}_1(f,x_1)$ and $\tilde{h}_2(f,x_2)$ belonging to these two families are parametrized by $x_1$ and $x_2$,
respectively.  In the main body of this paper, the two families $I_1$ and $I_2$ are the waveforms generated by
binaries in spin-orbit resonances with $\Delta\Phi=0^\circ$ and $\Delta\Phi = \pm180^\circ$.  Both families are
parametrized by the projected effective spin $\xi$ [Eq.~(\ref{eq:defxi})].  The overlap
\begin{equation}
{\cal O}(x_1, x_2) \equiv  {\cal O}[\tilde{h}_1(f,x_1), \tilde{h}_2(f,x_2)]
\end{equation}
defined by Eq.~(\ref{overlapdefinition}) induces two different mappings $F:I_1\rightarrow I_2$ and $G:I_2\rightarrow I_1$
between these families.  $F(x_1)$ is the member of $I_2$ with the highest overlap with $x_1$, and, conversely, $G(x_2)$
is the member of $I_1$ with the highest overlap with $x_2$.  The two mappings $F$ and $G$ for the resonant families
are shown by the two curves in Fig.~\ref{xi_xi_max}; both mappings are one-to-one and monotonically increasing, and
${\cal O}_{\rm max}(x_1) \equiv {\cal O}[x_1, F(x_1)]$ is a slowly varying function of $x_1$.  In this Appendix, we argue
that mappings satisfying these mild conditions are an approximate period-2 symmetry ($F\simeq G^{-1}$).

Since ${\cal O}(x_1, x_2)$ is slowly varying, we can Taylor expand about the point [$x_0, F_0 \equiv F(x_0)$] to find
\begin{align}
{\cal O}(x_1, x_2) \simeq& {\cal O}(x_0, F_0) + \frac{\partial{\cal O}}{\partial x_1}(x_1 - x_0)
+ \frac{1}{2} \frac{\partial^2{\cal O}}{\partial x_1^2}(x_1 - x_0)^2 \notag \\
& + \frac{\partial^2{\cal O}}{\partial x_1\partial x_2}(x_1 - x_0)(x_2 - F_0) \notag \\
& + \frac{1}{2} \frac{\partial^2{\cal O}}{\partial x_2^2}(x_2 - F_0)^2 + \ldots
\end{align}
where all derivatives are evaluated at the point [$x_0, F_0$], and the ellipsis represents higher-order terms.  There is no
linear term proportional to $(x_2 - F_0)$ in this expansion, because the mapping $F$ is defined to maximize the overlap
as a function of $x_2$ ($\partial{\cal O}/\partial x_2 = 0$).  We now wish to find $G_0 \equiv G(F_0)$, the best match in
$I_1$ for $F_0 \in I_2$.  This point [$G_0, F_0$] is found by setting $\partial{\cal O}/\partial x_1 = 0$.  Differentiating the
above Taylor expansion of the overlap, we find
\begin{equation} \label{E:dOdx1}
\frac{\partial{\cal O}}{\partial x_1} \Big|_{[G_0, F_0]} \simeq \frac{\partial{\cal O}}{\partial x_1} +
\frac{\partial^2{\cal O}}{\partial x_1^2}(G_0 - x_0) + \ldots = 0~.
\end{equation}
However,
\begin{equation} \label{E:dadx1}
\frac{d{\cal O}_{\rm max}}{dx_1} = \frac{\partial{\cal O}}{\partial x_1} + \frac{\partial{\cal O}}{\partial x_2}
\frac{dF}{dx_1} = \frac{\partial{\cal O}}{\partial x_1}
\end{equation}
since $F$ was constructed such that $\partial{\cal O}/\partial x_2 = 0$ at [$x_0, F_0$].  Solving (\ref{E:dOdx1}) for $G_0$
and substituting (\ref{E:dadx1}), we find
\begin{equation} \label{E:G0}
G_0 \simeq x_0 - \frac{\partial{\cal O}}{\partial x_1} \left( \frac{\partial^2{\cal O}}{\partial x_1^2} \right)^{-1} =
x_0 - \frac{d{\cal O}_{\rm max}}{dx_1} \left( \frac{\partial^2{\cal O}}{\partial x_1^2} \right)^{-1}.
\end{equation}
The color scale in Fig.~\ref{xi_xi_max} indicates that $|d{\cal O}_{\rm max}/d\xi_1| \lesssim 0.06$, while the second
derivatives at the maxima in Figs.~\ref{xi_ov_notmax} and \ref{predictors} suggest $|\partial^2{\cal O}/\partial \xi_1^2|
\gtrsim 30$.  Then Eq.~(\ref{E:G0}) implies $|\xi - G[F(\xi)]| \lesssim 2 \times 10^{-3}$, which is below our numerical
accuracy.  We therefore conclude that the mappings $F$ and $G$ are indeed inverses of each other at our current level of
approximation.

\bibliography{overlap}
\end{document}